\documentclass[preprint,floatfix]{revtex4}

\pdfoutput=1

\usepackage{pdfpages}
\usepackage{graphicx} 
\usepackage{dcolumn} 
\usepackage{amsmath}
\usepackage{amssymb}
\usepackage{bm}
\usepackage{times}
\usepackage{ulem}
\usepackage[justification=raggedright]{caption}

  \usepackage{boondox-cal}

\usepackage{subfigure}
\usepackage{tabularx}
\usepackage{appendix}

\usepackage{amstext}
\usepackage{array}

\usepackage{bbm}
\usepackage{blkarray}

\definecolor{wine-stain}{rgb}{0.5,0,0} 
\definecolor{bblue}{rgb}{0,0.0,0.5} 

\usepackage[colorlinks=true
,urlcolor=blue
,anchorcolor=blue
,citecolor=blue 
,filecolor=blue
,linkcolor=wine-stain
,menucolor=blue
,pagecolor=blue
,linktocpage=true
,pdfproducer=medialab
,linktoc=all 
]{hyperref}

\usepackage{booktabs}  

\allowdisplaybreaks

\newcommand{\ncmd}{\newcommand}

\ncmd{\lt}{\left}
\ncmd{\rt}{\right}
\ncmd{\tr}[1]{~\mbox{tr}\lt\{ {#1}\rt\}}
\ncmd{\half}{\frac{1}{2}}
\ncmd{\eps}{\epsilon}
\ncmd{\veps}{\varepsilon}
\ncmd{\dgr}{\dagger}
\ncmd{\sig}{\sigma}
\ncmd{\gam}{\gamma}
\ncmd{\rtarw}{\rightarrow}
\ncmd{\Rt}{\Rightarrow}
\ncmd{\abs}[1]{\lt\cb{#1}\rt\cb}
\ncmd{\avg}[1]{\lt\lb{#1}\rt\rb}
\ncmd{\sgn}[1]{\mbox{sgn}\lt(#1\rt)}
\ncmd{\kap}{\kappa}
\ncmd{\wtil}[1]{\widetilde{#1}}
\ncmd{\thrfr}{\therefore}
\ncmd{\eq}[1]{Eq. \eqref{#1}}
\ncmd{\fig}[1]{Fig. \ref{#1}}
\ncmd{\ordr}[1]{\mathcal{O}\lt(#1\rt)}
\ncmd{\dsty}{\displaystyle}
\ncmd{\alert}[1]{\color{red}{#1}}
\ncmd{\mc}{\mathcal}
\ncmd{\mbf}[1]{\mathbf{#1}}
\ncmd{\Deriv}[2]{\frac{d{#1}}{d{#2}}}
\ncmd{\ParDeriv}[2]{\frac{\partial{#1}}{\partial{#2}}}
\ncmd{\step}[1]{\Theta\lt(#1\rt)}
\ncmd{\td}{\tilde} 
\ncmd{\what}{\widehat}

\ncmd{\hphi}{\hat \phi} 
\ncmd{\hpi}{\hat \pi} 
\ncmd{\hK}{\hat K} 
\ncmd{\hL}{\hat L} 
\ncmd{\hU}{\hat U} 
\ncmd{\hQ}{\hat Q} 

\ncmd{\sL}{ \sqrt{L} }

\ncmd{\bqa}{\begin{eqnarray}} 
\ncmd{\eqa}{\end{eqnarray}}

\ncmd{\nn}{\nonumber \\}
\ncmd{\comment}[1]{{\color{red}{#1}}}
\definecolor{new_color}{RGB}{50,155,0}

\ncmd{\vrho}{\varrho}

\ncmd{\qstar}{\mathcal{q}}
\ncmd{\sstar}{\mathcal{ s}}
\ncmd{\pcstar}{ \mathcal{p}_{c}}
\ncmd{\tcstar}{ \mathcal{t}_{c}}

\ncmd{\ha}{\frac{1}{2}}

\ncmd{\lb}{\big<}
\ncmd{\rb}{\big>}
\ncmd{\cb}{\big|}
\ncmd{\cH}{{\cal H}}
\ncmd{\lc}{l_c}
\ncmd{\Lc}{l_c}
\ncmd{\ta}{\tilde \alpha}

\ncmd{\bt}{\bar t}
\ncmd{\bp}{\bar p}

\ncmd{\btc}{\bar t_c}
\ncmd{\bpc}{\bar p_c}
\ncmd{\bs}{\bar s}
\ncmd{\bq}{\bar q}
\ncmd{\bU}{\bar U}
\ncmd{\bQ}{\bar Q}
\ncmd{\bW}{\bar W}
\ncmd{\by}{\bar y}

\ncmd{\rot}{\br_A, \br_B}
\ncmd{\rto}{\br_B, \br_A}

\ncmd{\PhiAi}{\Phi^A_{~~i}}
\ncmd{\PhiAj}{\Phi^A_{~~j}}
\ncmd{\PhiAk}{\Phi^A_{~~k}}
\ncmd{\PhiBi}{\Phi^B_{~~i}}
\ncmd{\PhiBj}{\Phi^B_{~~j}}
\ncmd{\PhiBk}{\Phi^B_{~~k}}

\ncmd{\PiiA}{\Pi^i_{~~A}}
\ncmd{\PijA}{\Pi^j_{~~A}}
\ncmd{\PikA}{\Pi^k_{~~A}}

\ncmd{\PiiB}{\Pi^i_{~~B}}
\ncmd{\PijB}{\Pi^j_{~~B}}
\ncmd{\PikB}{\Pi^k_{~~B}}

\ncmd{\hPhi}{\hat \Phi}
\ncmd{\hPi}{\hat \Pi}
\ncmd{\hG}{{\hat G}}
\ncmd{\hH}{\hat H}
\ncmd{\hJ}{\hat J}

\ncmd{\GLL}{$GL(L, \mathbb{R})$ }
\ncmd{\SLL}{$SL(L, \mathbb{R})$ }

\ncmd{\bk}{{\bf k}}
\ncmd{\br}{{\bf r}}

\ncmd{\qw}{q^{(w)}} 
\ncmd{\qo}{q^{(1)}} 
\ncmd{\qt}{q^{(2)}}

\ncmd{\tds}{S}
\ncmd{\tdt}{T }
\ncmd{\tdp}{P}
\ncmd{\tdX}{X}
\ncmd{\tdY}{Y}

\ncmd{\hd}{\frac{r'}{2}}

\ncmd{\hPhiAi}{\hat \Phi^A_{~~i}}
\ncmd{\hPhiAj}{\hat \Phi^A_{~~j}}
\ncmd{\hPhiAk}{\hat \Phi^A_{~~k}}
\ncmd{\hPhiBi}{\hat \Phi^B_{~~i}}
\ncmd{\hPhiBj}{\hat \Phi^B_{~~j}}
\ncmd{\hPhiBk}{\hat \Phi^B_{~~k}}

\ncmd{\hPiiA}{\hat \Pi^i_{~~A}}
\ncmd{\hPijA}{\hat \Pi^j_{~~A}}
\ncmd{\hPikA}{\hat \Pi^k_{~~A}}
\ncmd{\hPiiB}{\hat \Pi^i_{~~B}}
\ncmd{\hPijB}{\hat \Pi^j_{~~B}}
\ncmd{\hPikB}{\hat \Pi^k_{~~B}}

\ncmd{\cA}{{\cal A}}
\ncmd{\cB}{{\cal B}}
\ncmd{\cC}{{\cal C}}
\ncmd{\cD}{{\cal D}}

\ncmd{\cG}{{\cal G}}
\ncmd{\cGI}{{\cal G}^{-1}}

\ncmd{\cGij}{{\cal G}^i_{~~j}}
\ncmd{\cGji}{{\cal G}^j_{~~iå}}

\ncmd{\cGIij}{({\cal G}^{-1})^i_{~~j}}
\ncmd{\cGIji}{({\cal G}^{-1})^j_{~~iå}}

\ncmd{\cstr}{\cb s, t_1, t_2 \rb}

\ncmd{\czr}{ \cb 0 \rb } 
\ncmd{\lzc}{ \lb 0 \cb }

\ncmd{\cepr}{ \cb q, \phi, \varphi \rb^{'} }

\ncmd{\tx}{\td x}
\ncmd{\ty}{\td y}

\ncmd{\cphir}{\cb \phi \rb}
\ncmd{\lphic}{\lb \phi \cb}

\ncmd{\ctr}{\cb T \rb}
\ncmd{\ltc}{\lb T \cb}
\ncmd{\cpr}{\cb p \rb}
\ncmd{\lpc}{\lb p \cb}

\ncmd{\cTr}{\cb {\cal T} \rb}

\ncmd{\ccr}{\cb \chi \rb}
\ncmd{\lcc}{\lb \chi \cb}

\ncmd{\cS}{{\cal S}}
\ncmd{\tcS}{\tilde { \cal S}}

\ncmd{\sqg}{\sqrt{|g(x)|}}
\ncmd{\gmn}{E_{\mu i}}

\ncmd{\emi}{E_{\mu i}}
\ncmd{\emj}{E_{\mu j}}
\ncmd{\eni}{E_{\nu i}}
\ncmd{\enj}{E_{\nu j}}
\ncmd{\de}{|E(x)|}

\ncmd{\gemn}{g_{E,\mu \nu}}
\ncmd{\gemnp}{g_{E',\mu \nu}}

\ncmd{\bgmn}{\bar E_{\mu i}}
\ncmd{\Fs}{F(\sigma)}
\ncmd{\cJ}{ {\cal J}\left(E',\sigma';E,\sigma \right)  }
\ncmd{\Si}{ \Sigma \left(E',\sigma';E,\sigma \right)  }

\ncmd{\grs}{g_{\alpha \beta}}
\ncmd{\pmn}{\pi^{\mu i}}
\ncmd{\prs}{\pi^{\alpha \beta}}
\ncmd{\fs}{ \frac{3}{\sqrt{2}} \sigma }
\ncmd{\fst}{ 3 \sqrt{2} \sigma }

\ncmd{\ee}{entanglement entropy }

\ncmd{\cT}{{\cal T} }
\ncmd{\cP}{{\cal P} }
\ncmd{\cV}{{\mathbcal V} }
\ncmd{\cW}{{\cal W} }

\ncmd{\ccw}{\cos( 2 \omega \delta) }
\ncmd{\ssw}{\sin (2 \omega \delta) }

\makeatletter
\newcommand*{\rom}[1]{\expandafter\@slowromancap\romannumeral #1@}
\makeatother

\begin{document}

\title{
Clock-dependent spacetime
}

\author{Sung-Sik Lee\\
\vspace{0.3cm}
{\normalsize{Department of Physics $\&$ Astronomy, McMaster University, Hamilton ON, Canada}}
\vspace{0.2cm}\\
{\normalsize{Perimeter Institute for Theoretical Physics, Waterloo ON, Canada}}
}

\date{\today}

\begin{abstract}

Einstein's theory of general relativity is based on the premise 
that the physical laws take the same form in all coordinate systems.
However, it still presumes 
a preferred decomposition of the total kinematic Hilbert space
into local kinematic Hilbert spaces.
In this paper, we consider a theory of quantum gravity 
that does not come with a preferred partitioning of the kinematic Hilbert space.
It is pointed out that, in such a theory,
dimension, signature, topology and geometry of spacetime 
 depend on how a collection of local clocks is chosen 
  within the kinematic Hilbert space.

\end{abstract}

\maketitle

\newpage

{
\hypersetup{linkcolor=bblue}
\hypersetup{
    colorlinks,
    citecolor=black,
    filecolor=black,
    linkcolor=black,
    urlcolor=black
}
\tableofcontents
}


\newpage

\section{Introduction}

In Newtonian paradigm, 
physical laws govern the evolution of dynamical degrees of freedom
with respect to one universal time.
Einstein's theory of relativity demotes time 
from the absolute status in two ways. 
First, the notion of simultaneity becomes observer-dependent
for events that are spatially separated,
and there is no universal sense of  past, present and future.
Second, time evolution is turned into a gauge transformation,
and time as a gauge parameter has no physical meaning by itself\cite{PhysRev.160.1113,1992gr.qc....10011I,1992grra.conf..211K,https://doi.org/10.1002/andp.201200147}.
One needs to use dynamical variables as clocks
to describe the relative evolution of other degrees of freedom\cite{PhysRevD.65.124013,Dittrich2007,Gambini_2004}. 
The theory only predicts correlations among dynamical degrees of freedom.

In quantum gravity,
choosing clocks boils down to dividing the kinematic Hilbert space
into a sub-Hilbert space for clocks and the rest for the `true' physical degrees of freedom.
In general relativity, while there is no preferred coordinate system, 
there is still a preferred way of
identifying a sub-Hilbert space for each local clock. 
This is because the theory is covariant only under the transformations
that preserve the integrity of local sites.
Under diffeomorphism, points in space are permuted, 
but the quantum information stored at a point is never spread over multiple points.
The preferred set of local kinematic Hilbert spaces is invariant under diffeomorphism,
and each local clock variable is chosen from one of the local Hilbert spaces.

Requiring that physical laws are covariant only under the 
local Hilbert space-preserving transformations may be too restrictive
in quantum gravity that has no predetermined notion of locality.
In priori, one partitioning of Hilbert space is no better than others.
Furthermore, the fact that locality is a dynamical concept in quantum gravity
obscures the distinction between local and non-local transformations. 
Consider a unitary transformation that mixes local Hilbert spaces 
associated with multiple sites.
For states in which those sites are within a short-distance cutoff scale,
the transformation can be regarded as local in space,
and we may gauge it as an internal symmetry.
However, it is no longer local for other states 
in which the sites affected by the transformation 
are macroscopically apart.
Once we gauge general unitary transformations
that do not preserve local kinematic Hilbert spaces,
the preferred Hilbert space decomposition is lost.

In this paper, we examine
consequences of having no preferred
Hilbert space decomposition
in a recently proposed model of quantum gravity\cite{Lee:2020aa}.
In the model, geometric degrees of freedom are  
collective variables of underlying quantum matter\cite{
Maldacena:1997re,Witten:1998qj,Gubser:1998bc}.
In the absence of a predetermined Hilbert space decomposition,
there exists a greater freedom in how clock variables are identified 
within the kinematic Hilbert space.
Instead of choosing local clock variables from predetermined local Hilbert spaces,
in this theory local Hilbert spaces are identified from a choice of clock variables.
In other words, the notion of local site is derived from clocks.
A set of local observers who use a particular set of local clocks 
constructs a geometry based on the pattern of entanglement
present across the local Hilbert spaces associated with the clocks.
Inasmuch as patterns of entanglement
depend on partitioning of the total  Hilbert space,
one state can exhibit different geometries with different choices of local clocks.
The spacetime that emerges with respect to one choice of clocks
is in general different from the spacetime that arises 
with respect to another set of local clocks.
The purpose of the paper is to show that 
the spacetimes that emerge from different choices of local clocks
can exhibit different dimensions, signatures, topologies and geometries.

The rest of the paper is organized as follows.
In Sec. \ref{sec:rev}, 
we review the theory introduced in Ref. \cite{Lee:2020aa}
as it forms the basis of the present work.
In the review, 
the gauge symmetry,
the constraint algebra,
and
the way an emergent geometry is identified 
from the underlying microscopic degrees of freedom 
are emphasized as they are the key ingredients needed in this paper. 
Sec. \ref{sec:frame} is the main part of the present paper.
The objectives of this section are two-folded. 
The first is to identify a set of clock variables 
to construct spacetime from the correlation 
between the clock variables and the remaining
physical degrees of freedom in the semi-classical limit.
The second is to examine how different choices of 
local clocks leads to different spacetimes.
In Sec. \ref{subsec:gaugeinvariant}, 
a procedure that generates gauge invariant states 
from a set of first-class constraints 
is discussed.
In Sec. \ref{subsec:constraint}, the gauge constraints
are solved to identify the constraint surface
in the semi-classical limit.
Sec. \ref{subsec:gaugefixing} discusses a gauge fixing prescription
that introduces clock variables 
and the associated local Hilbert spaces.
The section also discusses how 
the correlation between the remaining physical degrees of freedom 
and the local clocks determines an emergent spacetime.
In Sec. \ref{subsec:more}, two schemes
that use different sets of local clocks
are applied to one physical state,
and extract the spacetimes that emerge
from those choices. 
Sec. \ref{sec:summary} is a summary with open questions.

\section{Review : a model of quantum gravity with emergent spacetime }
\label{sec:rev}

In this section, 
we review the model introduced in Ref. \cite{Lee:2020aa}.
We recap the main results that are needed in this paper,
and refer the readers to the original paper for details.

\begin{itemize}
\item Kinematic Hilbert space

The fundamental degree of freedom is  a real rectangular matrix
with $M$ rows and $L$ columns with $M \gg L \gg 1$ : 
$\Phi^A_{~~i}$
with $A=1,2,..,M$ and $i=1,2,..,L$.
The row ($A$) labels  flavour,
and the column ($i$) labels sites.
The full kinematic Hilbert space ${\mathbb H}$ 
is spanned by the set of basis states, 
$\Big\{ \cb \Phi \rb  \Big| 
-\infty < \Phi^A_{~~i} < \infty ~\mbox{with}~
1 \leq A \leq M, 1 \leq i \leq L
\Big\}$,
where $ \cb \Phi \rb 
\equiv \otimes_{i,A} \cb \Phi^A_{~~i} \rb$,
and $ \cb \Phi^A_{~~i} \rb$ is the eigenstate of $\hat \Phi^A_{~~i}$. 
The inner product between basis states is 
$
\lb \Phi' \cb \Phi \rb =
 \prod_{i,A} \delta \left(
\Phi_{~~i}^{'A} - \Phi_{~~i}^A
\right)$.

\item Flavour (Global) symmetry

The global symmetry is $O(M)$.
It rotates the flavour index acting as a left multiplication on $\Phi$ :
$\Phi \rightarrow O \Phi$, where $O \in O(M)$. 
The generator of the $O(M)$ flavour symmetry is 
$\hat T_{\td o} = \frac{1}{2} \tr{ \left( \hPhi \hPi  -  \hPi^T \hPhi^T   \right) \td o}$, 
where
$\hPhi$ represents the operator valued $M \times L$  matrix,
$\hPi$  is the conjugate momentum that is an $L \times M$ matrix,
and
$\td o$ is a real $M \times M$ anti-symmetric matrix.

\item Frame

A frame is a decomposition of the total kinematic Hilbert space
into a direct product of local Hilbert spaces.
For example, ${\mathbb H}$ can be written as 
\bqa
{\mathbb H}= \otimes_i {\mathbb H}_i,
\label{eq:frame1}
\eqa
where ${\mathbb H}_i$ is the local Hilbert space 
spanned by  $\left\{ \otimes_A \cb \Phi^A_{~~i} \rb \right\}$ at site $i$.
The frame can be rotated with \SLL transformations
that act as right multiplications on $\Phi$ :
$\Phi \rightarrow \Phi g$,
where $g \in SL(L, \mathbb{R})$.
New basis states defined by
$\cb \Phi \big)   \equiv  \cb \Phi g \rb$
have the same inner product, 
$\big( \Phi \cb \Phi' \big)  = 
 \prod_{I,A} \delta \left(
\Phi_{~~I}^{'A} - \Phi_{~~I}^A
\right)$.
This allows us to write
$\cb \Phi \big)  = \otimes_{A,I} \cb \Phi^A_{~~I} \big)$.
The Hilbert space ${\mathbb H}'_I$ spanned by 
$\left\{ \otimes_A \cb \Phi^A_{~~I} \big) \right\}$
forms a local Hilbert space for site $I$
in the rotated frame,
and 
the kinematic Hilbert space can be decomposed as 
\bqa
{\mathbb H}= \otimes_I {\mathbb H}'_I.
\label{eq:frame2}
\eqa
In general, a state that is unentangled in one frame
has non-trivial inter-site entanglement in another frame.

\item Gauge symmetry

In the limit that the size of matrix becomes large, 
the sites can collectively form a space manifold.
We identify the emergent geometric degrees of freedom
from the microscopic degree of freedom
based on the algebra that gauge constraints obey.
Just as the momentum and Hamiltonian constraints
generate spacetime diffeomorphism in general relativity,
the present theory comes with
generalized momentum and Hamiltonian constraints.

\begin{enumerate}

\item Generalized momentum

The \SLL group that rotates frame is taken 
as the gauge group that generalizes the spatial diffeomorphism
in the general relativity.
The generalized momentum operator that generates \SLL is 
\bqa
\hat G_y = \tr{ \hat G y},
\label{Gy}
\eqa
where $\hG$ is an operator valued rank $2$ traceless tensor
 given by
$\hat G^i_{~j} =
 \frac{1}{2} \left( \hPiiA \hPhiAj + \hPhiAj \hPiiA \right)
- \frac{1}{2L} \left( \hPikA \hPhiAk + \hPhiAk \hPikA \right) \delta^i_j
$,
and 
$y$ is a traceless  $L \times L$ real matrix called the shift tensor.
Under the \SLL transformation, $\hPhi$ and $\hPi$ 
transform as
$e^{- i \hG_y }~ \hPhi   ~e^{ i \hG_y } 
=\hPhi g_y$
and 
$e^{- i \hG_y }~ \hPi   ~e^{ i \hG_y } 
=g_y^{-1} \hPi$,
where $g_y = e^{-y}$.

\item Generalized Hamiltonian

The Hamiltonian constraint is written as 
\bqa
\hat H_v =   \tr{ \hH v},
\label{Hx}
\eqa
where $\hH$ is an operator valued  rank $2$ symmetric tensor given by
$\hH^{ij} = \frac{1}{2} \left[
		\left(  - \hPi \hPi^T + \frac{ \ta }{M^2}  \hPi \hPi^T  \hPhi^T  \hPhi \hPi \hPi^T \right)^{ij} 
		+
		\left(  - \hPi \hPi^T + \frac{ \ta }{M^2}  \hPi \hPi^T  \hPhi^T  \hPhi \hPi \hPi^T \right)^{ji} \right]$,
$\tilde \alpha$ is a constant parameter of the theory, and
$v$ is an  $L \times L$ real symmetric matrix called the lapse tensor.
Under the \SLL transformation, $v$ transforms as 
$e^{- i \hG_y }~ \hat H_v  ~e^{ i \hG_y } 
=\hat H_{v'}$, 
where
$v'=  (g_y^{-1})^T  v  g_y^{-1 } $
with $g_y = e^{-y}$.

\item
Constraint Algebra

The generalized momentum and Hamiltonian constraints satisfy
the  first-class quantum algebra :
\bqa
 \left[  \hG^i_{~j}, \hG^k_{~l} \right] & = & i A^{ikn}_{jlm}~ \hG^m_{~n}, \nn
 \left[ \hG^i_{~j}, \hH^{kl} \right] & = &  i B^{ikl}_{jmn}~ \hH^{mn},  \nn
 \left[  \hH^{ij}, \hH^{kl} \right] & = & 
i \hat C^{ijkl n}_{m}~ \hG^m_{~n}
+ \frac{i}{M}\hat D^{ijkl }_{nm} ~ \hH^{mn},
\label{eq:constraintalgebra}
\eqa 
where
\bqa
 A^{ikn}_{jlm} & = & \delta^k_j \delta^i_m \delta^n_l - \delta^i_l \delta^k_m \delta^n_j, \nn
 B^{ikl}_{jmn} & = & \delta^k_j \delta^{il}_{mn} +  \delta^l_j \delta^{ki}_{mn}, \nn
\hat C^{ijkln}_{m} & = &
-4 \ta
\Bigl[
\hat U^{n(j} \hat U^{i)[l} \delta^{k]}_m - \hat U^{n [l} \hat U^{k](j} \delta^{i)}_m 
\Bigr]  \nn
&& + 4 \ta^2
 \Bigl[
-( \hU \hQ)^{(j}_{~[m} \hU^{\{ l [n} \hU^{n'] k \}} \delta^{i)}_{m']}
+ \hU^{ (j [n} ( \hQ \hU)_{[m'}^{~\{ k} \hU^{n'] i)} \delta^{l\}}_{m]} \nn
&& \hspace{1.2cm} +
 \hU^{ (j [n} ( \hU \hQ)^{\{ l}_{~[m} \hU^{n'] i)} \delta^{k\}}_{m']} 
- \hU^{ \{l [n} \hU^{n'] k\}} ( \hQ \hU)_{[m'}^{~( i}  \delta^{j)}_{m]} \nn
&&   \hspace{1.2cm} 
+  \frac{1}{M^2} \Bigl( 
M \hU^{(j [n} \hU^{n'] \{k} \delta^{l\}}_{[m} \delta^{i)}_{m']}
+ (M+2)  \hU^{(j [n} \hU^{\{l i) } \delta^{n']}_{[m} \delta^{k\}}_{m']} \nn
&&   \hspace{2.5cm} 
+ 2 \hU^{(j [n} \hU^{n'] i)} \delta^{\{l}_{[m} \delta^{k\}}_{m']}
- 2  \hU^{(j \{k} \hU^{l\} [n } \delta^{n']}_{[m} \delta^{i)}_{m']} \nn
&&   \hspace{2.5cm} 
- 2 \hU^{(j \{k} \hU^{[n' n]} \delta^{l\}}_{[m} \delta^{i)}_{m']}
- 2  \hU^{(ij) } \hU^{\{k [n } \delta^{n']}_{[m} \delta^{l\}}_{m']} 
\Bigr)
\Bigr] \delta^{m'}_{n'}, \nn
\hat D^{ijkl}_{nm} & = & 
-4i \ta \left( \hat U^{kl} \delta^{ij}_{nm} - \hat U^{ij} \delta^{kl}_{nm}  \right)
\label{eq:ABC}
\eqa
with $\hat U = \frac{1}{M}( \hPi \hPi^T)$,
 $\hat Q = \frac{1}{M}(\hPhi^T \hPhi)$
 and
$\delta_{ij}^{kl} = \frac{1}{2}
\left(
   \delta_{i}^{k} \delta_{j}^{l} 
+ \delta_{i}^{l} \delta_{j}^{k}  \right)$. 
While  $A^{ikn}_{jlm}$ and $B^{ikl}_{jmn}$ are constant tensors,
  $\hat C^{ijkln}_{m} $ and $\hat D^{ijkl}_{nm} $ are operator valued
  structure function.
In the expression for 
$\hat C^{ijkln}_{m}$,
each pair of indices in
$(i,j)$, $[n,n']$, $[m,m']$, $\{k,l\}$ are 
understood to be symmetrized.
In the large $M$ limit, 
$A^{ikn}_{jlm}, B^{ikl}_{jmn}, \hat C^{ijkln}_{m},\hat D^{ijkl}_{nm}   \sim O(1)$,
and the term that is proportional to $\hH^{mn}$ 
in $ \left[  \hH^{ij}, \hH^{kl} \right]$ is sub-leading.
\eq{eq:constraintalgebra} is the exact commutator
that remains well-defined at the quantum level.
In the large $M$ limit, 
the algebra 
reduces to the one that includes 
the hypersurface embedding algebra of the general relativity.
This plays the key role in identifying the emergent geometry
in this model. 

\end{enumerate}

\item
Emergent geometry

\begin{figure}[ht]
\begin{center}
\centering
\includegraphics[scale=0.5]{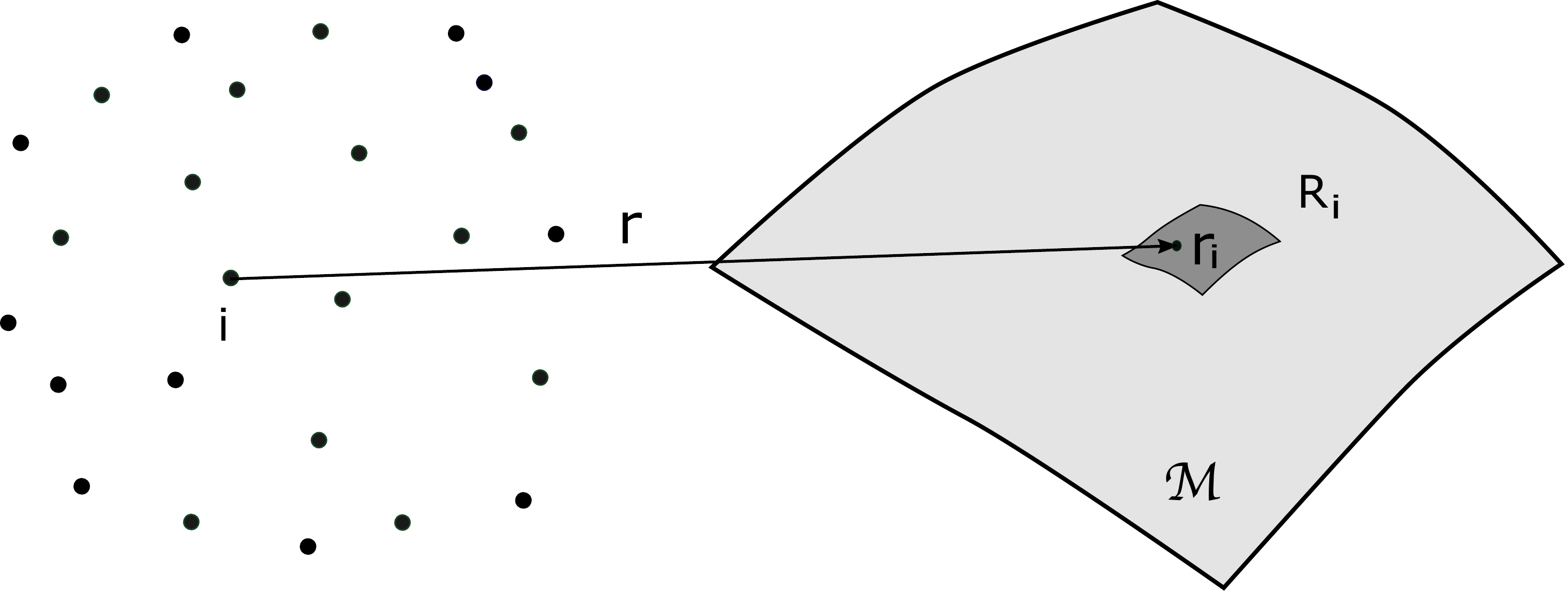}
\end{center}
\caption{
A coordinate system is a mapping 
from the set of sites in a frame into a manifold ${\cal M}$. 
$r_i$ represents the image of site $i$,
and $R_i$ is a simply connected region allocated to site $i$ such that 
$r_i \in R_i$, ${\cal M} = \cup_i R_i$ and $R_i \cap R_j = \emptyset$ for $i\neq j$. 
}
\label{fig:coordinate}
\end{figure}

A coordinate system is 
a mapping  $r : i \rightarrow r_i$ 
from the set of sites in a frame
to a manifold ${\cal M}$
with a region $R_i \ni r_i$ 
assigned to site $i$ (see \fig{fig:coordinate}).
In the large $L$ limit in which the image of sites is dense in  ${\cal M}$,
\eq{eq:constraintalgebra} induces an algebra on ${\cal M}$. 
The generators of the induced algebra include 
the Weyl generator ($\cD$),
the momentum density ($\cP_\mu$),
the Hamiltonian density ($\cH$)
and generators for higher spin gauge transformations.
Since $v$ transforms as a rank $2$ symmetric tensor under \SLL,
one can always choose a frame in which $v$ is diagonal.
In this frame,
the Hamiltonian constraint can be written as
\bqa
\hH_v & = & \int dr ~ 
   \cH(r)   \theta_v(r).
\label{eq:GyHv1}
\eqa
Here
$ \cH(r_i)  =  V_i^{-1}  \hH^{ii}$
is the Hamiltonian density,
$V_i$ is the coordinate volume of $R_i$, and
 $\theta_v(r_i)  =    v_{ii}$
 is the lapse function.
$\hG^i_{~j}$ is  now viewed as 
a bi-local operator defined on ${\cal M}$.
It can be expanded in the relative coordinate  as
\bqa
  \hG^i_{~j} =
  \hG^i_{~i}
  +
  \left.  \frac{\partial \hG^i_{~j}}{\partial r^\mu_j} \right|_{j=i} (r^\mu_j - r^\mu_i) + ...,
\eqa
where the ellipsis represents higher derivative terms.
This leads to
\bqa
\hG_y & = & \int dr ~\Bigl(
\cD(r) \zeta_y(r) +  \cP_\mu(r)  \xi_y^\mu(r) + .. 
\Bigr), 
\label{eq:GyHv2}
\eqa
where the derivative expansion is shown to the first order 
that includes the momentum density.
Here 
$\cD(r_i)  =  V_i^{-1} \hG^i_{~i}$ and
$\cP_\mu(r_i)  =   V_i^{-1}  \left.  \frac{\partial \hG^i_{~j}}{\partial r^\mu_j} \right|_{j=i}$
are the Weyl generator and the momentum density, respectively.
$\zeta_y(r_i)  =  \sum_j y^j_{~i}$ and
$\xi_y^\mu(r_i) =  \sum_j y^j_{~i}( r^\mu_j - r^\mu_i )$
represent
the Weyl parameter and the shift vector, respectively.

The commutators between $\cD$, $\cP_\mu$ and $\cH$
are completely determined from \eq{eq:constraintalgebra}.
To the leading order in 
 $1/M$ and the derivative expansion,
the  commutators read
 \bqa
&  \Bigl[ 
  \int dr ~\zeta_1(r) \cD(r), 
  \int dr ~\zeta_2(r) \cD(r) 
    \Bigr]&   =  0, \nn
& \Bigl[ 
  \int dr ~\xi^\mu(r) \cP_\mu(r), 
 \int dr ~\zeta(r) \cD(r)    \Bigr]  & =  
  i   \int dr ~  {\cal L}_{ \xi} \zeta(r)  ~ \cD(r), \nn
&  \Bigl[  \int dr ~\zeta(r) \cD(r) ,  \int dr ~\theta(r) \cH(r)  \Bigr] & =  
2 i  \int dr ~ \zeta(r) \theta(r) \cH(r), \nn   
& \Bigl[ 
  \int dr ~\xi_1^\mu(r) \cP_\mu(r), 
   \int dr ~\xi_2^\nu(r) \cP_\nu(r)
    \Bigr]  & =  
     i \int dr ~  {\cal L}_{ \xi_1} \xi_2^\mu(r)  ~ \cP_\mu(r), \nn
 & \Bigl[  \int dr ~\xi^\mu(r) \cP_\mu(r),
 \int dr ~\theta(r) \cH(r) \Bigr]  & =  
 i \int dr ~{\cal L}_{\xi} \theta(r)  ~ \cH(r), \nn
 & \Bigl[  \int dr ~\theta_1(r) \cH(r) ,  \int dr ~\theta_2(r) \cH(r)  \Bigr] & =  
  i  \int dr ~
   \left(  
    \hat F^\nu(r) \cD(r) 
+
   \hat G^{\mu \nu} 
       \cP_\mu(r)
   \right) 
   \left( \theta _1 \nabla_\nu \theta _2  - \theta _2 \nabla_\nu \theta _1 \right),
     \nn
\label{eq:hypersurface}
 \eqa 
 where 
 ${\cal L}_{\xi}$ represents the Lie derivative with respect to the vector field $\xi$,
 and
 $\hat F^\nu(r_m) $ and $\hat G^{\mu \nu}(r_m)$ are given by
\bqa
\hat F^\nu(r_m) & = & \frac{1}{2} \sum_{i,k,n} \hat C^{ i i k k n }_{m} \left( r^\nu_k - r^\nu_i \right), 
\label{Fnu} \\
\hat G^{\mu \nu}(r_m) & = & \frac{1}{2} \sum_{i,k,n} 
\hat C^{ i i k k n }_{m}  \left( r^\mu_n - r^\mu_m \right) \left( r^\nu_k - r^\nu_i \right).
\label{Gmunu2}
\eqa
The momentum and Hamiltonian densities obey an algebra 
that generalizes the hyper-surface deformation algebra of the general relativity\cite{PhysRev.116.1322,TEITELBOIM1973542}, 
provided that the metric is identified as the symmetric part
of $\hat G^{\mu \nu}$, 
 \bqa
\hat g^{\mu \nu}(r_m)  & = &
-
\frac{{\cal S}}{4}
 \sum_{i,k,n} \hat C^{ i i k k n }_{m} 
\Bigl[
 \left( r^\mu_n - r^\mu_m \right) \left( r^\nu_k - r^\nu_i \right)
+
  \left( r^\nu_n - r^\nu_m \right) \left( r^\mu_k - r^\mu_i \right)
 \Bigr],
\label{gmunu2}
\eqa
where ${\cal S}$ is the signature of the spacetime direction
translated by the Hamiltonian constraint.
The overall sign of the spacetime metric can be chosen either way.
In the rest of the paper, we choose the convention in which ${\cal S}=-1$.
$\hat F^\nu(r)$  and the anti-symmetric part of $\hat G^{\mu \nu}$
represent additional collective fields 
that generalize the hyper-surface deformation algebra of general relativity.

The contravariant metric in \eq{gmunu2}
is given by a second moment of $\hat C^{ i i k k n }_{m}$,
which measures a multi-point correlation in the system.   
If the range of entanglement and correlation is large in the coordinate distance,
the second moment becomes large accordingly, 
which results in a small proper distance between points in space.
The metric identified from the constraint algebra naturally captures 
the intuition that two sites that are strongly entangled
are physically close\cite{PhysRevLett.96.181602,
1126-6708-2007-07-062,
VanRaamsdonk:2010pw,
Lewkowycz2013,
Headrick:2014aa,
Faulkner:2013aa,
Lashkari:2014aa,
2013arXiv1309.6282Q,
2014JHEP...03..051F,PhysRevD.95.024031}.
On the other hand, the metric captures only a specific pattern of entanglement,
and there also exist non-geometric entanglements.
For example, there exist finely tuned states
in which two points that are infinitely far still have
$O(M)$ entanglement through other channels such as 
the higher-order moments of $\hat C^{ i i k k n }_{m}$
\cite{Lee:2020aa}.
In this sense, EPR is strictly `bigger' than ER in the present theory\cite{doi:10.1002/prop.201300020}.

States for which there exist coordinate systems with well-defined metric 
in the large  $M$ and $L$ limit
form a special set of states,
and are referred to have local structures.
For a state with a classical local structure,  
there exists a coordinate system  
associated with a well-defined manifold such that
$\lb \hat g^{\mu \nu } \rb \equiv \frac{\lb \Psi \cb \hat g^{\mu \nu} \cb \Psi \rb }{\lb \Psi \cb \Psi \rb}$  is 
invertible and smooth on the manifold,
and
$ \lb  \left( \hat g^{\mu \nu} - \lb \hat g^{\mu \nu} \rb \right)^2  \rb
 \rightarrow 0 $ in the large $M$ and $L$ limit.
The dimension, topology and geometry of the manifold are properties of state.

$\hH_v$ in \eq{Hx} is a non-local Hamiltonian as a quantum operator,
but it is relatively local\cite{Lee2018}
in the following sense\footnote{
We note that this is different from the relative locality 
introduced in Ref. \cite{PhysRevD.84.084010}.}.
Suppose $\cb \Psi \rb$ has a local structure in a frame.
To this state, $\hH_v$ with a lapse tensor diagonal in that frame
acts as a local Hamiltonian 
to the leading order in the large $M$ limit, that is,
\bqa
\hH_v \cb \Psi \rb \approx \hat H_{eff}^\Psi \cb \Psi \rb,
\label{eq:HvHpsi}
\eqa
where $\hat H_{eff}^\Psi$ is a Hamiltonian
that is local in the manifold associated 
with the local structure of $\cb \Psi \rb$.
The discrepancy between 
$\hH_v $ and $\hat H_{eff}^\Psi$
in \eq{eq:HvHpsi} is 
sub-leading  in $1/M$.
This can be understood by writing
the Hamiltonian with the lapse tensor $v=I$ as 
\bqa
\hat H_v
 &\approx&
 -  \hPi^{i}_A \hPi^{i}_A
+
\frac{ \ta }{M^2} 
 \lb \hPi^{i}_{~A} \hPi^{j}_{~A} \rb
   \hPhi^{B}_{~~j}  \hPhi^{B}_{~~k}
 \lb \hPi^{k}_{~A} \hPi^{i}_{~A} \rb,
\eqa
where all repeated indices are summed over.
In the large $M$ limit, the fluctuation of 
$\hPi^{i}_{~A} \hPi^{j}_{~A}$ is small,
and the replacement of the operator with its expectation 
value is valid to the leading order in the large $M$ limit.
The second term in the Hamiltonian can be viewed
as a hopping term between sites $j$ and $k$
whose hopping amplitude is proportional to 
the expectation value of $ \lb \hPi^{i}_{~A} \hPi^{j}_{~A} \rb
  \lb \hPi^{k}_{~A} \hPi^{i}_{~A} \rb$ in a state.
  If the two-point function $\lb \hPi^{i}_{~A} \hPi^{j}_{~A} \rb$
  is short-ranged as a function of $r_i - r_j$ in a manifold, 
  the Hamiltonian effectively
  behaves as a local Hamiltonian in the manifold.
 The Hamiltonian acts in a state-dependent manner
to the leading order in the large $M$ limit,
and the local properties of the effective Hamiltonian 
are inherited from the state\cite{Lee2019}.



\end{itemize}

\section{Clocks and emergent spacetime}
\label{sec:frame}

\subsection{Gauge invariant states}
\label{subsec:gaugeinvariant}

The physical Hilbert space is given by 
the set of gauge invariant
states that satisfy
\bqa
\hat H_v \cb \chi \rb = 0, ~~
\hat G_y \cb \chi \rb = 0
\eqa 
for any lapse tensor $v$ and shift tensor $y$.
A gauge invariant state can be constructed by
projecting an arbitrary state 
to the physical Hilbert space.
The projection can be implemented with 
a series of gauge transformations applied to
an initial trial state $\cb \chi \rb$ as
\bqa
\cb 0_\chi \rb =
\lim_{Z \rightarrow \infty}
\int {\cal D}v  \int {\cal D}y ~ 
e^{-i \varepsilon \left(  \hat H_{v^{(1)}} + \hat G_{ y^{(1)}} \right) } 
e^{-i \varepsilon \left( \hat H_{v^{(2)}} + \hat G_{ y^{(2)}} \right) } 
...
e^{-i \varepsilon \left( \hat H_{v^{(Z)}} + \hat G_{ y^{(Z)}} \right) } 
\cb \chi \rb,
\label{eq:project} 
\eqa  
where $\varepsilon$ is a non-zero constant,
 $\int {\cal D} v \equiv \int \prod_{l=1}^Z D v^{(l)}$
 and 
 $\int {\cal D} y \equiv \int \prod_{l=1}^Z D y^{(l)}$
denote the sum over all possible combinations of the lapse and shift tensors\footnote{
While \eq{eq:project} is equivalent to the state obtained 
by one projection,
$\int Dv Dy ~
e^{-i \left(  \hat H_{v} + \hat G_{y} \right) } 
\cb \chi \rb$,
\eq{eq:project} is more convenient to use 
in the path integral formalism 
by taking small $\varepsilon$ limit.}.
The resulting state is gauge invariant
if it does not vanish
(see appendix  \ref{eq:appgaugeinv}).

Although the momentum and Hamiltonian constraints 
are invariant under the $O(M)$ flavour symmetry, 
a gauge invariant state may spontaneously 
break the global symmetry to a smaller group.
To simplify the problem of 
extracting the dynamical information from gauge invariant states,
it is convenient to focus on a sector 
with a definite flavour symmetry group.
Let us denote the set of all states (gauge invariant or not) 
that respect the global symmetry $\Gamma \subset O(M)$ as $\cV_\Gamma$.
Basis states of $\cV_\Gamma$
can be labeled by a set of collective variables.
The bigger $\Gamma$ is, 
the less collective variables are needed to span $\cV_\Gamma$.
If $\Gamma$ is too big, 
there are too few kinematic collective variables to 
support non-trivial physical degrees of freedom
after the gauge degrees of freedom are removed.
One simple choice of $\Gamma$ that supports 
a minimal number of non-trivial physical degrees of freedom
is  $\Gamma^* = S_{L}^f \times O(N/2) \times O(N/2)$
with $N =M-L$\cite{Lee:2020aa}.
Here $S_L^f$ is the permutation group acting on the first $L$ flavours.
Two $O(N/2)$ groups generate flavour rotations
within the remaining two sets of $N/2$ flavours.
Basis states for $\cV_{\Gamma^*}$ can be written as
\bqa
\cstr & = & 
\sum_{P \in S_L^f }
\int D \Phi ~ e^{  i  \left[
\sqrt{N} \sum_{a,a'=1}^L 
 s^i_{~a} P^a_{a'} \Phi^{a'}_{~i} 
+ 
\sum_{b=L+1}^{L+\frac{N}{2}} 
 t_1^{ij} 
\Phi^b_{~i} \Phi^b_{~j}
+
\sum_{c=L+\frac{N}{2}+1}^{L+N}
 t_2^{ij} 
\Phi^c_{~i} \Phi^c_{~j}
\right]
  }  
    \cb \Phi \rb, 
\label{eq:symm}
\eqa
where $s, t_1, t_2$ are collective variables that label the basis states;
 $s$ is $L \times L$ matrix
and $t_1, t_2$ are $L \times L$ symmetric matrices.
In \eq{eq:symm}, all repeated site indices ($i,j$) 
are understood to be summed over from $1$ to $L$.
Due to the sum over the flavour permutations $P \in S_L^f $ in \eq{eq:symm},
$\cb s, t_1, t_2 \rb = \cb s P, t_1, t_2 \rb$ for $P \in S_L^f$.
%

$s$ transforms as $s \rightarrow g s$ under $SL(L, \mathbb{R})$,
and as $s \rightarrow s O$ under $O(L) \subset O(M)$,
where $g \in SL(L, \mathbb{R})$ and $O \in O(L)$.
An invertible $s$ breaks \SLL down to the subgroup,
${\cal I} = \{ s P s^{-1}  \cb   P \in S_L^f ~\mbox{with}~ \det P =1 \}$.
This follows from $( s P s^{-1}  ) s = s P \sim s$.
The unbroken gauge group is related to the even site-permutation group through 
a similarity transformation.
Therefore, $s$ acts as a Stueckelberg field
that breaks the generalized spatial diffeomorphism
to the discrete permutation group.
On the other hand, $t^{ij}_1$ and $t^{ij}_2$ are bi-local fields
that generate inter-site entanglement.
The mutual information between sites $i$ and $j$
for state in \eq{eq:symm} is proportional to 
$-N \sum_c 
\frac{|t^{ij}_c|^2}{Im t^{ii}_c Im t^{jj}_c } 
\ln \frac{|t^{ij}_c|^2}{Im t^{ii}_c Im t^{jj}_c } $
to the leading order in the small $t^{ij}_c$ limit\cite{Lee2016}.
Geometry is determined from the connectivity formed
by these bi-local fields.
Generic choices of $t^{ij}_c$ would break \SLL completely.
If $t^{ij}_c$ depends only on $r_i - r_j$ in a coordinate system,
the global translational symmetry in the manifold remains unbroken. 

General states in $\cV_{\Gamma^*}$ can 
be written as
\bqa
\ccr  = \int D s Dt_1 Dt_2  ~ \cstr  
\chi(s,t_1,t_2),
\label{ccr}
\eqa
where $ \chi(s,t_1,t_2)$ is a wavefunction of the collective variables.
In \eq{ccr}, the integrations over $s, t_1, t_2$ are defined
along the real axis of each component of the matrices.
If we choose the initial state $\cb \chi \rb$ from $\cV_{\Gamma^*}$,
it follows that $\cb 0_\chi \rb \in \cV_{\Gamma^*}$ in \eq{eq:project} 
because  $\hG$ and $\hH$ 
are invariant under the $O(M)$ favour rotation.
Therefore, \eq{eq:project} can be represented
as a path integration over $s, t_1, t_2$
and their conjugate variables.
By taking the small $\varepsilon$ limit after the large $Z$ limit is taken first,
\eq{eq:project} can be written as 
\bqa
\cb 0_\chi \rb & = & 
\int
 D s^{(0)} Dt^{(0)} 
 \int {\cal D} s {\cal D} t {\cal D} q {\cal D} p
 {\cal D} v {\cal D} y
~
\cb s^{(\infty)}, t_1^{(\infty)},   t_2^{(\infty)} \rb
e^{ i S } 
~
\chi( s^{(0)}, t_1^{(0)}, t_2^{(0)}  ). 
\label{eq:pathint}
\eqa
Here $S$ is the action for the collective variables and their conjugate momenta,
\bqa
S & = & 
N \int_0^{\infty}  d \tau ~
\mbox{tr} \Biggl\{
- q \partial_{\tau}  s
-  p_c \partial_{\tau} t_c 
- v(\tau) \cH[q(\tau),s(\tau), p_1(\tau),t_1(\tau),   p_2(\tau),t_2(\tau) ] \nn
&& 
\hspace{5cm} - y(\tau) \cG[q(\tau),s(\tau), p_1(\tau),t_1(\tau),   p_2(\tau),t_2(\tau) ]
\Biggr\}. \label{Sfinal} 
\eqa
$\cH$ and $\cG$ are the induced Hamiltonian and momentum constraints, respectively,
\bqa
&& \cH[q,s,p_1,t_1,p_2,t_2]  = 
-U + \ta U Q U
+ O \left( \frac{1}{N} \right),  
 \label{H0}
\\
&& \cG[q,s,p_1,t_1,p_2,t_2]  =  
  \left( 
  s q  +  2 \sum_c  t_c p_c
  - i \frac{M}{2N}  I  \right).
 \label{G0}
 \eqa
Here
$U^{ij} =  \left( s s^{T} + \sum_{c=1}^2 \left[ 4  t_c p_c t_c - i t_c \right] \right)^{ij}$ and
$Q_{ij} = \left( q^T q   +  p_1+p_2 \right)_{ij}$.
 $q$ is a $L \times L$ matrix that is conjugate to $s$.
$p_1$ and $p_2$ are symmetric $L \times L$ matrices conjugate to $t_1$ and $t_2$, respectively.
While $s$, $t_1$ and $t_2$ represent the `sources', 
the conjugate variables represent the corresponding
operators, $q^a_{~~i} = \frac{1}{\sqrt{N}} \Phi^a_{~~i}$
with $1 \leq a \leq L$,
 $p_{1,ij} = \frac{1}{N} \sum_{b=L+1}^{L+N/2} \Phi^b_{~~i} \Phi^b_{~~j}$ and
 $p_{2,ij} = \frac{1}{N} \sum_{c=L+N/2+1}^M  \Phi^c_{~~i} \Phi^c_{~~j}$\cite{Lee:2020aa}.
In total, there are $D_k = 2 L^2 + 2 L (L+1)$ kinematic phase space variables.
 ${\cal D} x \equiv \prod_{l=1}^\infty D x^{(l)}$
and $x(\tau) = x^{(l)}$ with $\tau = l \varepsilon$
for $x = s, q, t_c, p_c, v, y$.
$\tau$ is the parameter that labels 
the evolution of dynamical variables along gauge orbits.

All gauge invariant states 
have an infinite norm with respect to the inner product 
of the underlying Hilbert space.
The non-normalizability of gauge invariant states is attributed to the fact
that gauge orbits defined in the infinite-dimensional kinematic Hilbert space
are non-compact\cite{Lee:2020aa}. 
This is fine because
the dynamical variables include 
both clocks and physical
degrees of freedom,
and a gauge invariant state encodes the information
about an entire spacetime history.
In the large $N$ limit with $L >>1$,
the path integration in \eq{eq:pathint} is well approximated 
by the saddle-point approximation.
In this paper, we  study the classical dynamics of the theory 
 in the semi-classical limit.
In particular, we identify a set of local clocks 
from the dynamical variables, 
and construct a spacetime 
from the correlation between
the clocks and the remaining dynamical variables.
We will see that different choices of local clocks
lead to different spacetimes.

\subsection{Constraint surface}
\label{subsec:constraint}

From now on, we denote the saddle-point configuration
as  $\{ q, s, t_1, t_2, p_1, p_2 \}$,
using the same collective variables that appear in the path integration.
As an initial state in \eq{eq:project},
we consider a semi-classical state in which 
both the collective variables and their conjugate momenta
are well defined.
An example is the gaussian wavepacket considered in Ref. \cite{Lee:2020aa}.
Let us denote a semi-classical state 
whose collective variables are peaked at 
$\{  q,s,p_1,t_1,p_2,t_2 \}$
as $\cb \Psi_{ q,s,p_1,t_1,p_2,t_2 } \rb$.
Because of the permutation symmetry $S^f_L$ in $\Gamma^*$,
$\cb \Psi_{ q,s,p_1,t_1,p_2,t_2 } \rb = \cb \Psi_{  Pq, sP^T ,p_1,t_1,p_2,t_2 } \rb$
for any $P \in S^f_L$.
To the leading order in $1/N$, 
the application of the gauge transformation results in
\bqa
e^{-i \varepsilon \left(  \hat H_{v} + \hat G_{y} \right) } 
\cb \Psi_{ q,s,p_1,t_1,p_2,t_2 } \rb
\approx
e^{
-i \varepsilon N 
\tr{ \cH[q,s,p_1,t_1,p_2,t_2] v + \cG[q,s,p_1,t_1,p_2,t_2] y }
} 
\cb \Psi_{ q',s',p_1',t_1',p_2',t_2'} \rb,
\label{eq:semiproject}
\eqa
where
$x'=
x + \varepsilon \Bigl\{  x, \tr{ \cG y} \Bigr\}_{PB} +  \varepsilon \Bigl\{  x, \tr{ \cH v} \Bigr\}_{PB}$
for $x = \{ q, s, t_1, t_2, p_1, p_2 \}$,
and
$\{ A, B \}_{PB}  = 
\left(
    \frac{\partial A}{\partial q^\alpha_{~i } }  \frac{\partial B}{\partial s^i_{~\alpha} } 
-
\frac{\partial A}{\partial s^i_{~\alpha} }  \frac{\partial B}{\partial q^\alpha_{~i } }
      \right)
+ 
\delta^{kl}_{ij}
\left(
  \frac{\partial A}{\partial p_{c, ij}} \frac{\partial B}{\partial t_c^{kl} }
- \frac{\partial A}{\partial t_c^{kl} }  \frac{\partial B}{\partial p_{c,ij}}
\right)$ is the Poisson bracket.
In the large $N$ limit,
the semi-classical initial state survives the projections in \eq{eq:project} 
only if the collective variables and conjugate momenta 
satisfy the momentum and Hamiltonian constraints {\it classically},
\bqa
\tr{ \left(   s q  +  2 \sum_c  \td t_c p_c \right) y } & = & 0,
 \label{G} \\
\tr{ \left(
-U + \ta U Q U \right) v } &=& 0   \label{H}
 \eqa
for arbitrary traceless matrix $y$ (shift tensor)
and symmetric matrix $v$ (lapse tensor)\footnote{
Otherwise, the 
fast phase oscillation in \eq{eq:semiproject}
results in the destructive interference 
upon integrating over $v$ and $y$.
}.
Here 
$\td t_c = t_c - \frac{i}{8 p_c}$
in terms of which $U$ is written as
$U = \left( s s^T + \sum_c \left[ 4 \td t_c p_c \td t_c + \frac{1}{16} p_c^{-1} \right] \right)$.
Eqs. (\ref{G}) and (\ref{H}) give rise to $D_c = (L^2-1) + \frac{L(L+1)}{2}$ constraints.

Now we solve these constraints
to remove $D_c$ kinematic variables. 
The momentum constraint 
in  \eq{G} 
is readily solved by expressing $s$ in terms of $\td t_c$, $p_c$ and $q$ as
\bqa
s = \left( \beta I -   2 \sum_c \td t_c p_c \right)  q^{-1}
\label{eq:cons} 
\eqa
for any constant  $\beta$.
For $U \neq 0$, the Hamiltonian constraint in \eq{H} is equivalent to
\bqa
 \alpha U Q & = & I.   
 \label{eq:con2}
\eqa
Plugging \eq{eq:cons} into \eq{eq:con2},
we obtain a quadratic matrix equation for $\td t_1$,
\bqa
\td t_1 A^2 \td t_1 + \td t_1 B + B^T \td t_1 + C = 0,
\label{eq:t1}
\eqa
where
$A = 2 \sqrt{ p_1 q^{-1} (q^{-1})^{T} p_1 + p_1 }$\footnote{
The square root of a symmetric matrix can be defined as follows.  
A real symmetric matrix $X$ can be written as 
$X = O_X D_X O_X^T$,
where $D_X$ is a diagonal matrix and $O_X$ is an orthogonal matrix.
Its square root is given by $\sqrt{X} = O_X D_X^{1/2} O_X^T$.
},
$B = 2 p_1 q^{-1} (q^{-1})^{T} ( 2 p_2 \td t_2 - \beta )$,
$C = ( 2 \td t_2 p_2 - \beta ) q^{-1} (q^{-1})^{T} ( 2 p_2 \td t_2   - \beta )
+ 4 \td t_2 p_2 \td t_2 + \frac{1}{16} ( p_1^{-1} + p_2^{-1} ) - \frac{1}{\td \alpha} Q^{-1}$.
The solution to \eq{eq:t1} is written as
\bqa
\td t_1 = 
- A^{-2} B + A^{-1} O \sqrt{B^T A^{-2} B - C},
\label{eq:t1solution}
\eqa
where $O$ is an orthogonal matrix that should be chosen
so that $\td t_1$ is symmetric.
For every orthogonal matrix $O$ that satisfies
\bqa
- A^{-2} B + A^{-1} O \sqrt{B^T A^{-2} B - C}
=
-B^T A^{-2} + \sqrt{B^T A^{-2} B - C} O^T A^{-1},
\label{eq:eqO}
\eqa
\eq{eq:t1solution} is a solution
to  \eq{eq:con2}.
In general, 
\eq{eq:eqO} admits a discrete set of solutions 
because it contains $\frac{L(L-1)}{2}$ equations 
with the same number of unknowns.
If $A, B, C$ can be simultaneously diagonalized,
$O= \mbox{diag} ( \pm1, \pm1, ....)$ 
gives the solutions
 in the diagonal basis.
At least locally in the phase space, 
the generalized momentum and Hamiltonian constraints are solved
by expressing $s$ and $\td t_1$ in terms of
$\{  \beta, q, p_1, \td t_2, p_2 \}$.
This results in the $(D_k- D_c)$-dimensional constraint surface
on which the gauge constraints are satisfied classically.

\subsection{Gauge fixing }
\label{subsec:gaugefixing}

  \begin{figure}[ht]
\begin{center}
\centering
\includegraphics[scale=0.5]{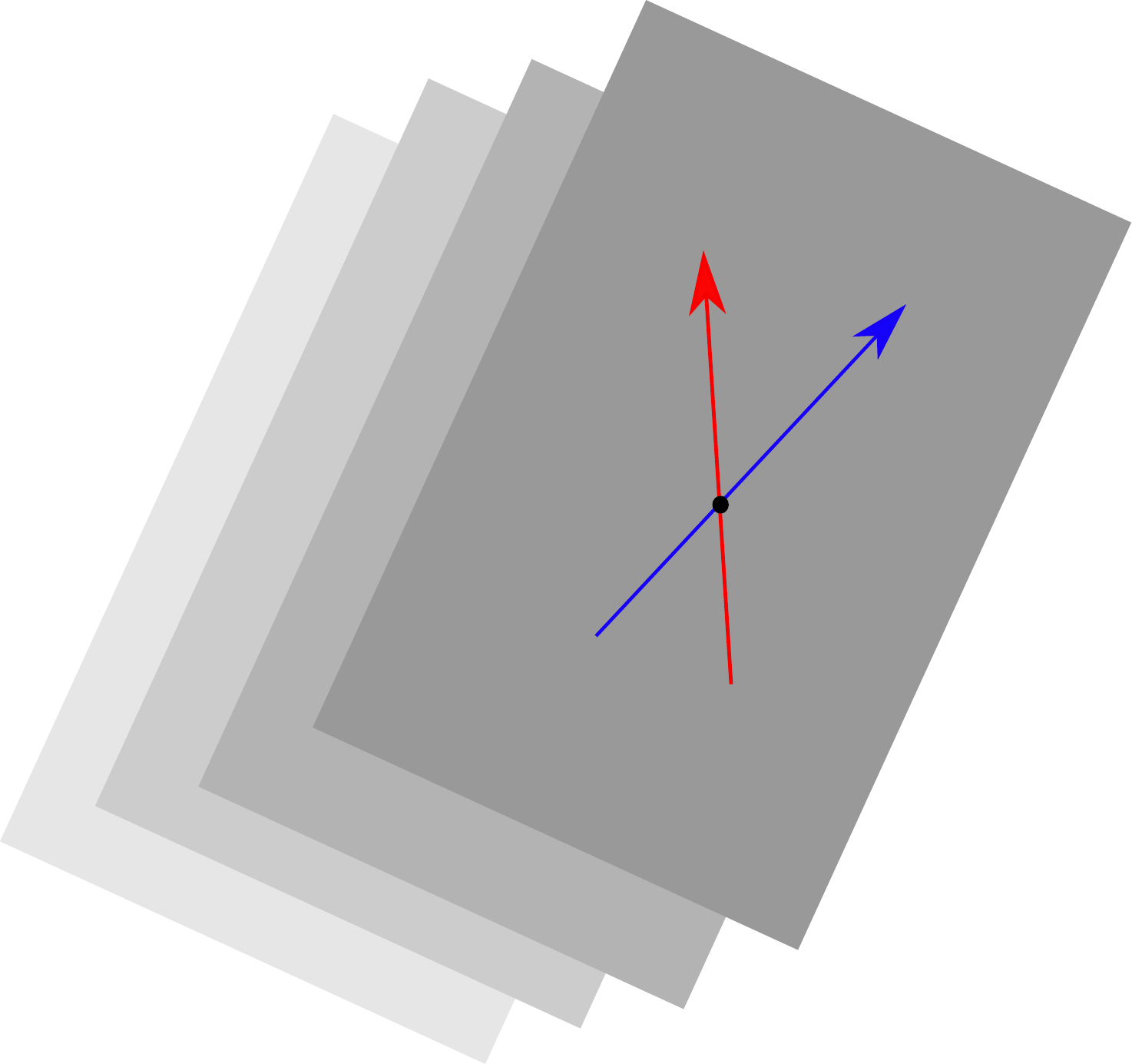}
\end{center}
\caption{
Each sheet represents a $D_c$-dimensional gauge manifold embedded
in the $(D_k - D_c)$ - dimensional constraint surface.
Any two configurations within a gauge manifold can be transformed to each other
through gauge transformations.
Different choices of the shift and lapse tensors generate different gauge orbits within a gauge manifold
as represented by arrows in the figure.
}
\label{fig:gaugeorbit}
\end{figure}

Because the constraints obey the first-class algebra,
the gauge orbits generared by $\cH$ and $\cG$ 
from an initial state in the constraint surface
remain within the constraint surface.
The equation of motion for the gauge orbit reads
\bqa
\partial_\tau \td t_c & = & -4 \td t_c v \td t_c   - \ta U v U  + \frac{1}{16} \frac{1}{p_c} v \frac{1}{p_c} - y \td t_c - \td t_c y^T, \nn
\partial_\tau p_c & = &  
4 p_c \td t_c v + 4 v \td t_c p_c   + p_c y + y^T p_c, \nn
\partial_\tau s & = & 
-2 \ta U v U q^T - y s, \nn
\partial_\tau q & = & 
  2 s^T v  + q y,
\label{eq:EOM3}
\eqa
where $y(\tau)$ and $v(\tau)$ are the shift and lapse tensors, respectively.
Because the shift and lapse tensors comprise $D_c$ independent gauge parameters,
the set of configurations generated from the gauge transformations
with all possible choices of $y$ and $v$
forms a $D_c$-dimensional {\it gauge manifold} 
(see \fig{fig:gaugeorbit}).
Configurations within a gauge manifold are physically equivalent,
and only $L(L+1)+2$  variables are left
to distinguish one gauge manifold from another.
These are the physical degrees of freedom.
To isolate the physical degrees of freedom,
we need to fix the gauge associated with the 
shift and lapse tensors.
This amounts to choosing a set of local clocks
and a coordinate system relative to which
the dynamical correlation of the remaining physical degrees of freedom
is expressed.

\begin{figure}[ht]
\begin{center}
\centering
\includegraphics[scale=0.5]{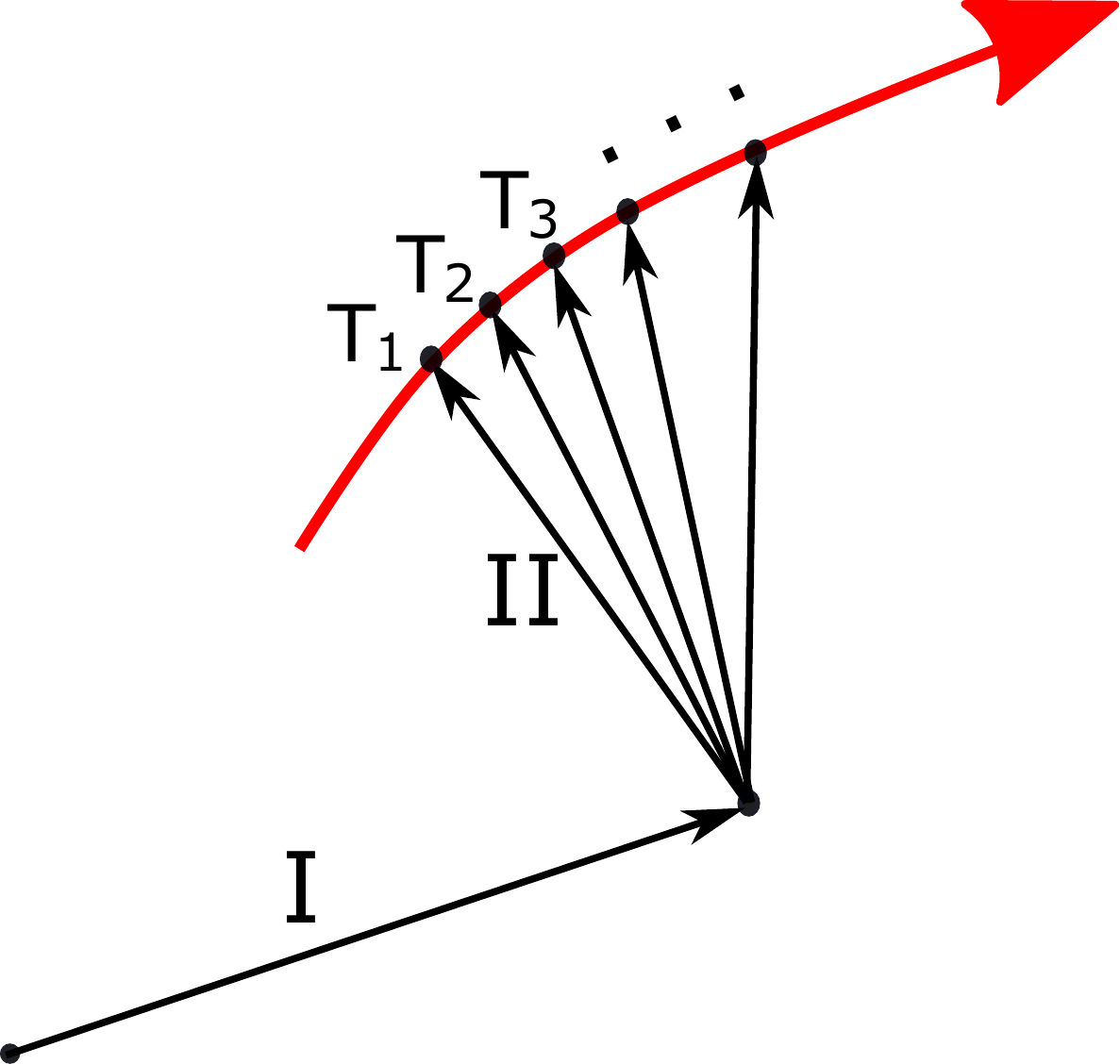}
\end{center}
\caption{
Two stages of gauge transformations
that bring an arbitrary initial state to a final state
that satisfies the gauge fixing condition in
\eq{eq:p2gauge}. 
In the first stage of gauge transformation (stage I),
an initial state is transformed so that the gauge fixing condition, 
$q= q_d g_f$ is enforced on $q$.
In the second stage of gauge transformation  (stage II),
the state obtained from the first gauge transformation is 
brought to the final form 
in \eq{eq:p2gauge}.
The one-parameter family of configurations,
denoted as the big red arrow,
with varying $p_d= T_1, T_2, ...$
describes a spacetime history,
where $p_d$ plays the role of time. 
}
\label{fig:gaugefixing}
\end{figure}

\subsubsection{Fixing the shift tensor}
\label{subsec:fixshift}

The momentum constraint  generates  \SLL transformations.
If \eq{eq:EOM3} is evolved for parameter time $\tau_1$
with $v=0$, one obtains
\bqa
\td t_c(\tau_1) & = & g(\tau_1) ^{-1} \td t_c(0) (g(\tau_1) ^{-1})^T, \nn
p_c(\tau_1) & = & g(\tau_1) ^{T} p_c(0) g(\tau_1) , \nn
s(\tau_1) & = & g(\tau_1) ^{-1} s(0), \nn
q(\tau_1) & = & q(0) g(\tau_1) ,
\label{eq:SLL}
\eqa
where $g(\tau_1)  = \bar {\cal P}_\tau e^{\int_0^{\tau_1} d\tau y(\tau)} \in$ \SLL.
$ \bar {\cal P}_\tau$ orders the matrix multiplication so that 
$e^{ d\tau y(\tau)}$ with smaller $\tau$ are placed 
to the left of the terms with larger $\tau$. 
For $q$ with $\det q \ge 0$,
the choice of
\bqa
g(\tau_1)= q_d(0) ~q(0)^{-1}  g_f,
\label{eq:sllfixing}
\eqa
with $q_d(0) = [\det q(0)]^{1/L}$ and  $g_f \in $ \SLL
leads to
\bqa
q(\tau_1) = q_d(0) ~ g_f.
\label{eq:sllgaugecondition}
\eqa
For a given $g_f$, \eq{eq:sllgaugecondition} completely fixes the gauge freedom associated with \SLL :
$g(\tau_1)$ in \eq{eq:sllfixing} is the only element in \SLL
that satisfies \eq{eq:sllgaugecondition}.
This gauge fixing amounts to locking site indices (columns)
with reference to the flavour indices (rows). 
We refer to the frame 
in which $q = q_d~ g_f$ 
as $g_f$-frame.
The path that connects an initial configuration
to the one that satisfies \eq{eq:sllgaugecondition}
is denoted as path I in \fig{fig:gaugefixing}.

\subsubsection{Fixing the lapse tensor}
\label{subsec:fixlapse}

In priori, there is no preferred frame, 
and any $g_f$ can be used in \eq{eq:sllgaugecondition}.
Here, we choose a frame in a clock-dependent way.
It is natural to use $p_1$ as our clock variables.
Because both $p_1$ and the lapse tensor
have the same number of variables,
the freedom associated with the lapse tensor 
can be fixed with a gauge condition imposed on $p_1$
up to a potential discrete ambiguity.
With $p_1$ chosen as the clock variable, 
selecting a particular $p_1$ 
along with \eq{eq:sllgaugecondition} 
corresponds to choosing a moment of time. 
Being a symmetric matrix, 
$p_1$ can be fixed 
with $\frac{L(L+1)}{2}$ gauge fixing conditions.
We take $L$ eigenvalues of $p_1$ 
as the readings of local clocks
defined at each site in a frame.
The other $\frac{L(L-1)}{2}$ components
encode the information on the frames in which $p_1$ is diagonal.
Since the clocks do not create inter-site entanglement in the frames 
in which $p_1$ is diagonal\footnote{
Because  $p_{1,ij} = \frac{1}{N} \sum_{b=L+1}^{L+N/2} \Phi^b_{~~i} \Phi^b_{~~j}$,
off-diagonal elements of $p_1$
generate inter-site entanglement.},
we choose $g_f$ in \eq{eq:sllgaugecondition}
so that $p_1$ is diagonal in the $g_f$-frame
\footnote{
There always exist frames in which $p_1$ is diagonal.
Suppose that  
$p_1 = X$ in the $g_f$-frame,
where $X$ is a general $L \times L$ symmetric matrix.
Under a frame rotation,
$ q' = q_d g_f g$ and
$ p_1' = g^T X g$,
where $g \in SL(L, \mathbb{R})$.
One can always choose $g$ such that $p_1' = p_d I$,
where  $p_d$ is a real number.
Now the clock takes the diagonal form 
in the $g_f'$-frame, where $g_f' = g_f g$.}.
Therefore, specifying a moment of time requires 
not only the readings of $L$ local clocks
but also the information on 
which part of the kinematic Hilbert space
is being used as $L$ local clocks.

Here, the clocks play dual roles.
First, the clocks provide a preferred frame dynamically. 
For different states of the clock,
we use different frames to decompose the 
total kinematic Hilbert space into local Hilbert spaces.
In this sense, the notion of local sites is provided by the clocks.
Second, the clocks provide a physical time
relative to which the evolution of other dynamical variables is tracked.
The correlation between $p_1$ and other degrees of freedom
describes the time evolution of the physical degrees of freedom
relative to the clocks.

 \begin{figure}[ht]
 \begin{center}
 \includegraphics[scale=0.35]{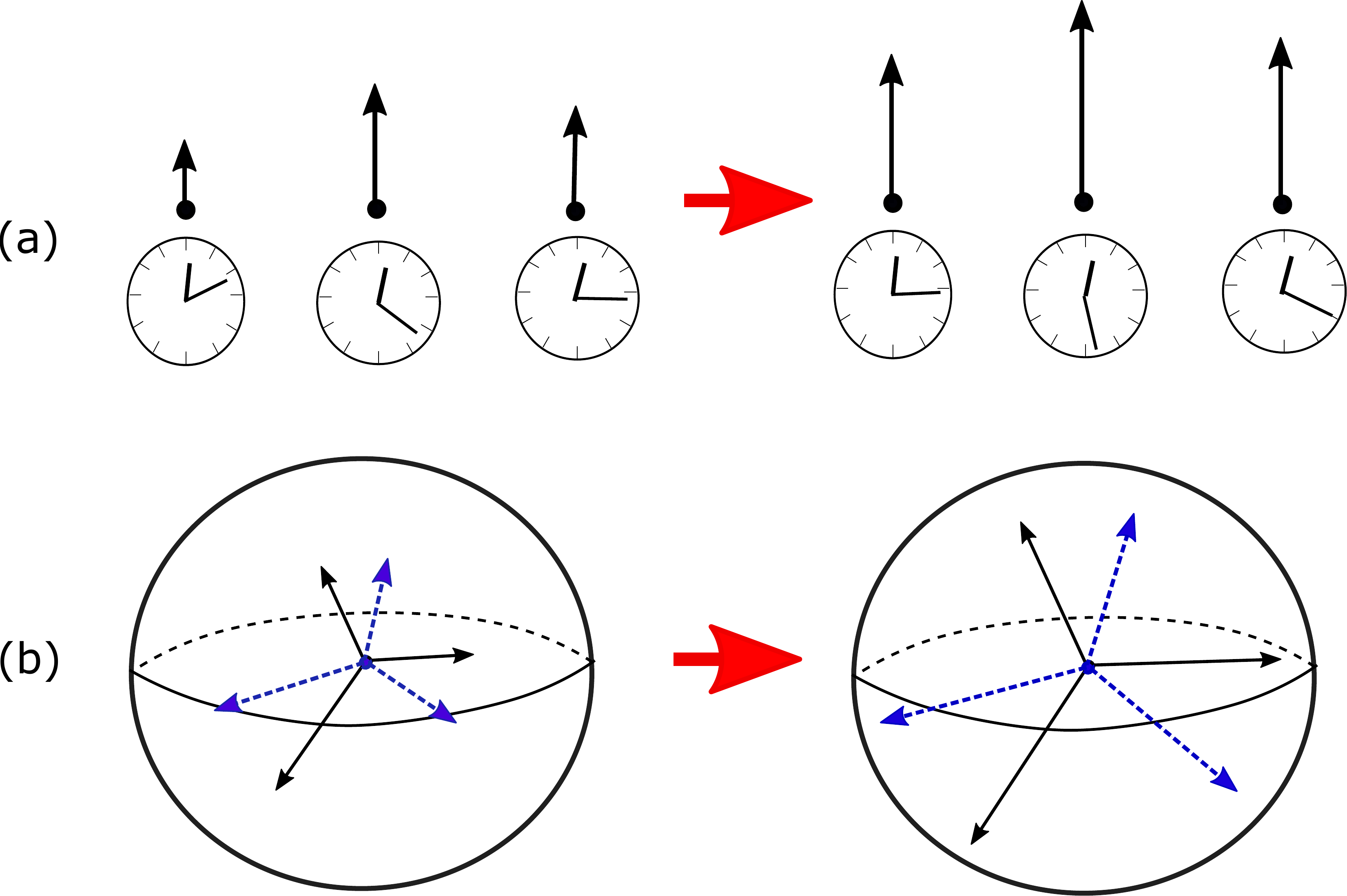} 
  \end{center}
 \caption{
 (a) In theories with a preferred set of local Hilbert spaces such as  general relativity,
 a moment of time is determined if a clock variable
 in each fixed local Hilbert space is specified.
 The length of the arrow at each site represents the time at that site.
The correlation between other physical degrees of freedom 
and the local clocks describes a time evolution 
from left to right in the figure.
 (b) In the present theory,  
the notion of local Hilbert spaces is determined by frame.
A frame is specified by $L$ vectors 
that form a parallelepiped,
where each vector represents a local site.
Once a frame is fixed, a local clock can be defined at each site.
	 In this figure, the set of solid (black) arrows represent 
a moment of time defined by a set of local clocks in one frame,
	 and the set of dashed (blue) arrows represent a moment of time
defined by local clocks in another frame.
The length of each arrow denotes the reading of the local clock at the corresponding site.
The spacetimes that emerge for different sets of local clocks
are in general different.
 }
  \label{fig:clocks}
 \end{figure}

It is instructive to compare the role of clocks 
in general relativity and the present theory. 
In general relativity, 
the four-dimensional spacetime can be sliced
into different stacks of three-dimensional spatial manifolds,
depending on the choice of the lapse function.
To specify a moment of time across the system, 
one has to fix the lapse function
by imposing a gauge fixing condition
on a scalar function in space.
The scalar function at each position in space 
plays the role of an internal clock at that position.
In the present theory, 
one needs to specify both frame and 
diagonal elements of $p_1$ 
in that frame to define a moment of time. 
In other words, one has to specify both the local 
Hilbert spaces and the readings of local clocks 
in the chosen local Hilbert spaces.
A moment of time chosen in one frame does not 
correspond to a moment of time in another frame 
unless $p_1$ takes diagonal forms in both frames.
This is illustrated in \fig{fig:clocks}.

The dynamical information of the theory is encoded in the correlation
between the clocks and the remaining physical degrees of freedom.
The state of physical degrees of freedom 
given as a function of state of the clocks 
is a prediction of the theory.
To extract this correlation,
we impose the following gauge fixing conditions on $q$ and $p_1$,
\bqa
q = q_d ~ g_f, ~~~ 
p_{1} = p_d~  I,
\label{eq:p2gauge}
\eqa
where $g_f \in$ \SLL 
and $q_d$, $p_d$ are real variables.
$p_{1,ii}$ serves as the local clock 
at site $i$ in the $g_f$-frame.
Here, the gauge is chosen so that all local clocks run uniformly.
In general, one could choose a non-uniform gauge condition such as
$p_{1,ij}=p_{d,i} \delta_{ij}$.
After the gauge fixing in \eq{eq:p2gauge},
what is left is the $L(L+1)+2$ physical degrees of freedom : 
$\{ p_2, \td t_2, q_d, \beta \}$. 
We now ask how the physical degrees of freedom evolve
as functions of $p_d$.
This describes the spacetime that emerges
for the set of clocks localized in the $g_f$-frame. 
Since $\beta$ is a constant of motion along the gauge orbit\cite{Lee:2020aa},
we will focus on the evolution of $p_2, \td t_2, q_d$.

From the discussion in Sec. \ref{subsec:fixshift},
we already know that the first condtion 
in  \eq{eq:p2gauge} can be readily imposed through an \SLL
transformation.
To impose the second condition in \eq{eq:p2gauge},
the configuration $\{ \td t_c(\tau_1), p_c(\tau_1), s(\tau_1), q(\tau_1) \}$ in \eq{eq:SLL}
with $q(\tau_1)=q_d(\tau_1) g_f$ and a generic $p_1(\tau_1)$ 
is further evolved with the equation of motion in \eq{eq:EOM3}.
During this evolution,
the shift and lapse tensors are chosen so that
$p_1$ at $\tau_2 > \tau_1$  satisfies 
\eq{eq:p2gauge}.
To make sure that  the gauge fixing condition for $q$ 
is maintained along the evolution, 
the shift is chosen to be
\bqa
y = - 2 q^{-1} s^T v + 2  \langle q^{-1} s^T v \rangle I,
\label{eq:shift}
\eqa
where $\langle A \rangle \equiv \frac{1}{L} \tr A$.
This guarantees that $q$ is proportional to $g_f$ 
at all $\tau$ irrespective of the lapse tensor.
To transform $p_{1}(\tau_1)$ to  the desired form of $p_1(\tau_2) = p_d I$,
we write the equation of motion for $p_1$ as
\bqa
\partial_\tau p_1 & = &  w p_1 + p_1 w^T,
\label{eq:pw}
 \eqa
where
$ w = v [ 4  \td t_1   - 2 s (q^{-1})^T ] +  2  \langle q^{-1} s^T v \rangle I $,
and  choose $v(\tau)$ such that
\bqa
w(\tau) p_1(\tau) + p_1(\tau) w^T(\tau)  = 2 p_d I - p_{1}(\tau_1) -  p_1(\tau)
\label{eq:eqforv}
\eqa
for $\tau_1 \leq \tau \leq \tau_2$.
\eq{eq:eqforv} is a set of $\frac{L(L+1)}{2}$ linear equations
for $v(\tau)$ at each $\tau$, and admits a unique solution in general.
It is straightforward to show that with this choice of the lapse tensor
$p_1( \tau_2 ) = p_d I$ at $\tau_2 = \tau_1 + \ln 2$.
This is denoted as path II in \fig{fig:gaugefixing}.

For a given initial state,
the physical variables obtained at $\tau_2$ depend on
$p_d$ and $g_f$. 
Therefore, the physical variables at $\tau_2$ can be written as
\bqa
\Bigl\{ p_2(p_d;g_f), \td t_2(p_d;g_f), q_d(p_d;g_f) \Bigr\}.
\label{eq:confgT}
\eqa 
Within the constraint surface, 
the spatial metric in \eq{gmunu2} is given by\cite{Lee:2020aa}
\bqa
g^{\mu \nu} 
&=&
 -2 \ta \sum_{l,n} U^{nl} U^{lm} 
 \left( r^\mu_{nm} r^\nu_{lm} + r^\nu_{nm} r^\mu_{lm} \right) 
\label{eq:gmunu4}
\eqa
with $r^{\mu}_{nm} = r^\mu_{n} - r^\mu_m$
to the leading order in $1/N$.
Consequently, \eq{eq:gmunu4} 
gives the spatial 
metric $g^{\mu \nu}(r, p_d; g_f)$ 
that depends on space ($r$) and time ($p_d$)
in the $g_f$-frame.
The correlation between the spatial metric and the physical clocks
describes a spacetime that emerges for the set of observers
who use local clocks chosen in the $g_f$-frame.

For some $p_d$,
there may be no lapse and shift tensors
that brings the initial state to the one 
that satisfy the gauge fixing condition in \eq{eq:p2gauge}.
It is also possible that 
a constant $p_1$ surface intersects with a gauge orbits 
multiple times.
In this case, $p_1$ can not be used
as a time variable globally\cite{1992gr.qc....10011I,1992grra.conf..211K}.
Here we don't attempt to find a global time variable.
We will be content with the fact
that $p_1$ serves as a set of clocks 
locally in the phase space.

\subsection{Multi-fingered internal time}
  \label{subsec:more}

 \begin{figure}[ht]
 \begin{center}
   \subfigure[]{
 \includegraphics[scale=0.35]{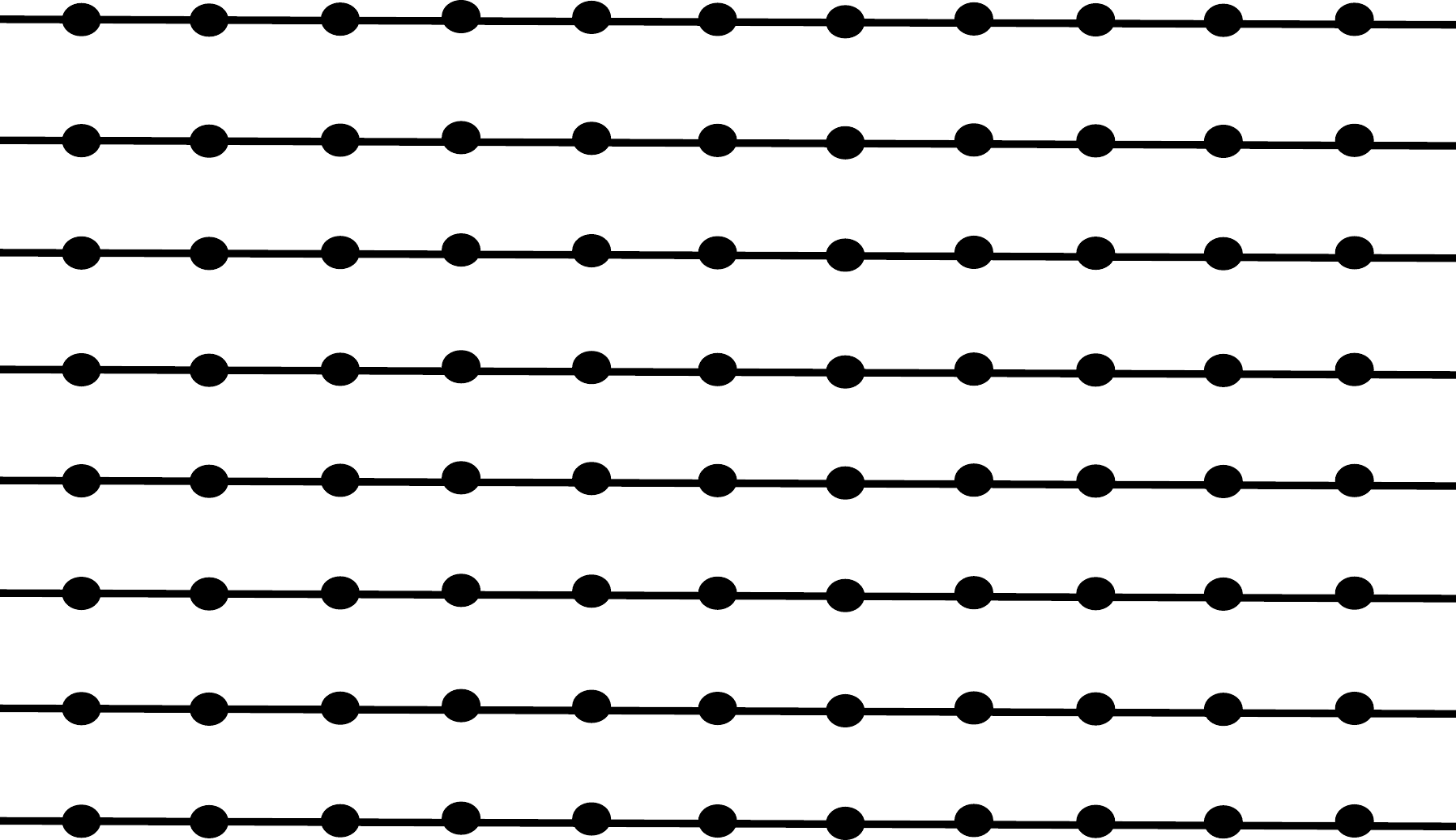} 
  \label{fig:2a}
 } 
  \hfill
 \subfigure[]{
 \includegraphics[scale=0.35]{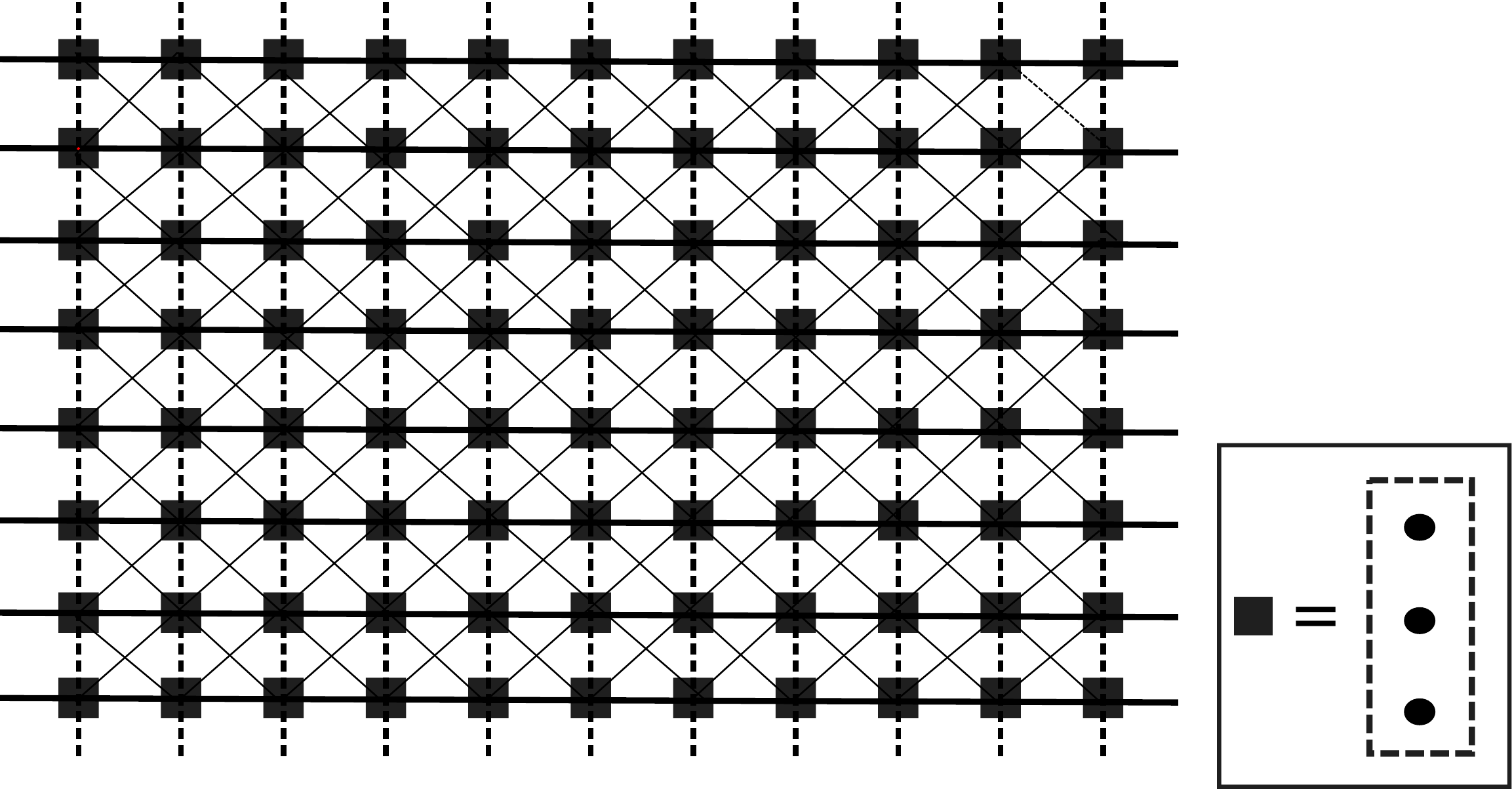} 
 \label{fig:2b}
 } 
  \end{center}
 \caption{
 (a) One-dimensional local structure of the state in \eq{eq:1d}.
 Each dot represents a site in the $I$-frame
 in which $q$ is proportional to the identity matrix.
  Links between sites represent
 non-zero collective variables ($p_{2,ij}$, $t_{2,ij}$)
 that create entanglement between the sites.
 According to \eq{gmunu2}, the state in \eq{eq:1d} gives rise to 
 $\sqrt{L}$ decoupled one-dimensional manifolds. 
 (b) In the $g_o$-frame, the state in \eq{eq:1d} is
 represented as \eq{eq:2d2}.
One site (represented as a square) in the $g_o$-frame 
 is composed of sites that belong to three different chains
 in the $I$-frame.
As a result, sites are entangled in both $x$ and $y$ directions
	 in the $g_o$-frame.
To avoid clutter in the figure, the links are drawn
only for the bi-local fields up to $O(\zeta)$.
For this state, \eq{gmunu2} gives a two-dimensional manifold.
 }
 \label{fig:graph_12}
 \end{figure}

The fact that one can choose any $g_f \in $ \SLL in \eq{eq:p2gauge}
encodes the freedom in choosing a frame 
in which local clocks are defined.
Under a rotation of frame,
a state in a local Hilbert space 
can be transformed to a linear superposition of 
states that belong to multiple local Hilbert spaces.
As a result, one state can exhibit different local structures 
in different frames.
To illustrate this through a concrete example, 
let us consider a semi-classical state with
\bqa 
p_{2,ij} = p_0  \delta_{ i_x, j_x } \delta_{ i_y, j_y }+ 
\epsilon 
\left( 
\delta_{ (i_x-j_x)_{\sL},1} 
+\delta_{ (i_x-j_x)_{\sL},-1} 
\right)
 \delta_{ i_y, j_y }, 
~~
\tilde t_{2} = 0, ~~
p_{1} = I,
~~
q =  I.
\label{eq:1d}
\eqa
Here, the site index $i$ with $i=1,2,..,L$ is labeled 
in terms of a pair of indices, $(i_x, i_y)$,
where $i_x, i_y = 1,2,..,\sqrt{L}$\footnote{
We assume that $L$ is the square of a whole number.
}.
$(x)_{\sL}  = x ~\mbox{mod} \sqrt{L}$.
$s$ and $\td t_1$ are determined from 
 Eqs. (\ref{eq:cons}) and (\ref{eq:t1solution}).
 Because \eq{eq:1d} has the translational invariance
 under $(i_x,i_y) \rightarrow (i_x+1, i_y)$ for each $i_y$,
all collective variables can be simultaneously diagonalized.
Among the possible solutions in  \eq{eq:t1solution},
we choose the branch with $O=+I$.
In  \eq{eq:1d}, the bi-local collective variable $p_2$ connects 
site $(i_x,i_y)$ with sites $(i_x \pm 1,i_y)$.
On the other hand, there is no entanglement between sites with different $i_y$
to the leading order in $1/M$.
Therefore, the state has an one-dimensional classical local structure.
It describes $\sL$ copies of one-dimensional manifold
with the periodic boundary condition as is shown in  \fig{fig:2a}.
This state breaks \SLL down to the discrete translation
$(i_x, i_y) \rightarrow (i_x+1, i_y)$
and the permutation group that interchanges $i_y$.

\subsubsection{Finger 1}
\label{subsubsection:finger1}

In this section, we consider the spacetime
that emerges for a set of local observers 
who use the diagonal elements of $p_1$ as local clocks in the $I$-frame.
This frame is defined by the gauge fixing conditions,
\bqa
q = q_d ~ I, ~~~ p_1 = p_d ~I.
\label{eq:p1gauge}
\eqa
Since \eq{eq:1d} satisfies  \eq{eq:p1gauge}
with $q_d=1$ and $p_d=1$,
\eq{eq:1d} is already on the desired gauge orbit. 
To move along the gauge orbit,
we take \eq{eq:1d} as the initial condition,
and evolve it with the lapse and shift tensors 
that maintain the gauge fixing conditions 
in \eq{eq:p1gauge} along the orbit.
We choose the lapse tensor,
\bqa
v =
 \left( 4  \td t_1   - 2 s (q^{-1})^T  \right)^{-1}
 \label{eq:lapsep1gauge}
\eqa
with the shift tensor given in \eq{eq:shift}.
With this choice, the equation of motion for $p_1$ becomes
\bqa
\partial_\tau p_1 = 
2  \Biggl[ I + 2 \left< q^{-1} s^T 
 \left( 4  \td t_1   - 2 s (q^{-1})^T  \right)^{-1}
\right> \Biggr] p_1,
\eqa
and \eq{eq:p1gauge} are satisfied along the trajectory.
The physical degrees of freedom are also evolved 
with the same lapse and shift tensors.
This gives the information on how the physical degrees of freedom 
are correlated with the clock variable $p_d$. 
The evolution results in the spacetime history
measured by the clocks that are local in the $I$-frame.

In the $I$-frame, it is convenient to introduce the one-dimensional coordinate system, $r_i = i_x$
for each decoupled ring.
In this coordinate system,
$p_{2, ij}$ and $\td t_2^{ij}$ in \eq{eq:1d} are short-ranged in $r_i-r_j$.
Since $\td t_1$ and $s$ are determined from $p_2, \td t_2, q$ from 
Eqs. (\ref{eq:cons}) and (\ref{eq:t1solution}),
$\td t_1^{ij}$ and $\left(s (q^{-1})^T \right)^{ij}$ also decays exponentially
in $r_i-r_j$ in the one-dimensional manifold.
Accordingly, the lapse tensor $v_{ij}$ in \eq{eq:lapsep1gauge} also decays exponentially
in $r_i-r_j$.
This guarantees that $\hH_v$ is relatively local in the one-dimensional coordinate system\cite{Lee:2020aa}.
Consequently, the decoupled chains remain decoupled 
under the Hamiltonian evolution  to the leading order in $1/M$.
Furthermore, the Hamiltonian acts as a local one-dimensional Hamiltonian
within each chain\cite{Lee2019}.
As a result, the emergent spacetime consists of $\sqrt{L}$ 
identical two-dimensional spacetimes that remain decoupled throughout the evolution.

 \begin{figure}[ht]
 \begin{center}
   \subfigure[]{
 \includegraphics[scale=0.8]{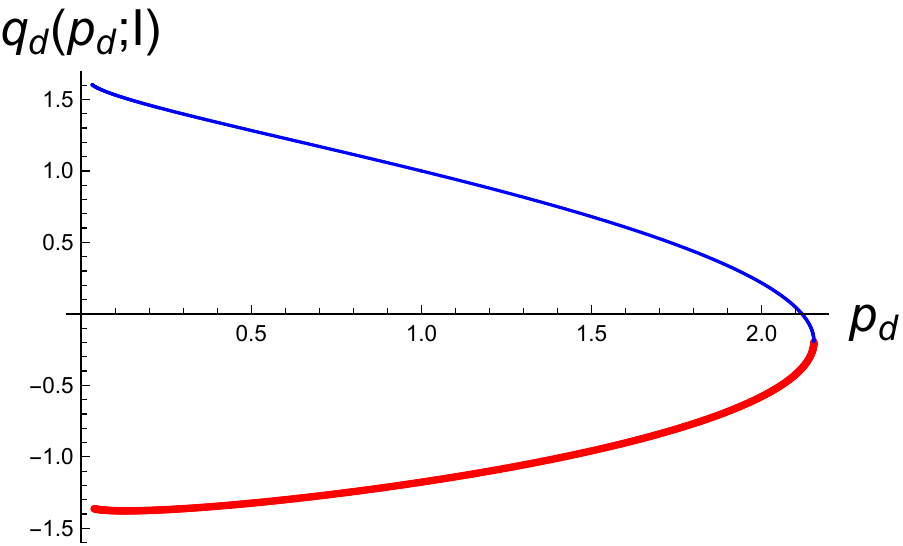} 
  \label{fig:q1p1d}
 } 
  \hfill
 \subfigure[]{
 \includegraphics[scale=0.8]{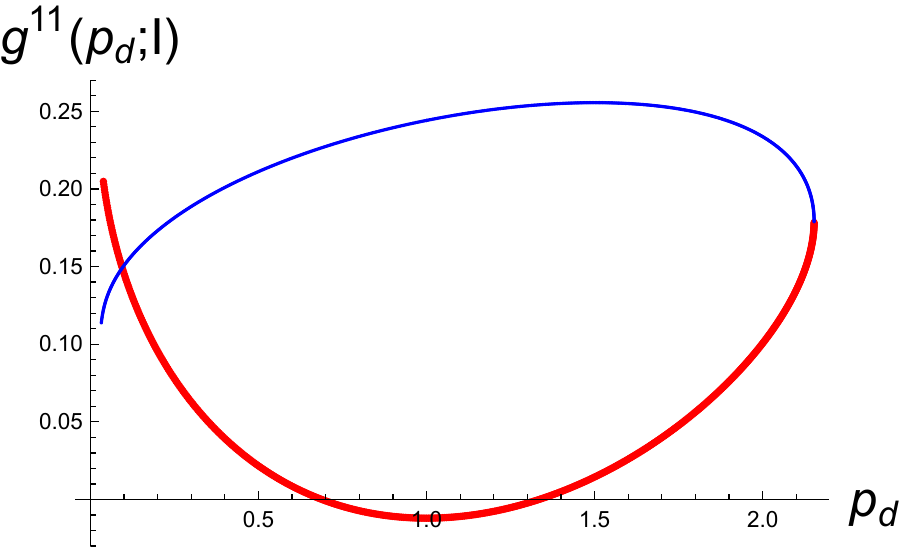} 
 \label{fig:g11p1_1d}
 } 
  \end{center}
 \caption{
  The evolution of $q_d$ and $g^{11}$
 as functions of $p_d$
 that emerge in the $I$-frame 
 for the initial condition given by \eq{eq:1d}
 with $p_0=1$, $\epsilon=0.2$
 for $\tilde \alpha=0.1$,  $\beta =0.2$ and $L=10^4$.
Different chains remain decoupled throughout the evolution,
and $g^{22} = 0$ at all time (not plotted here).
For each value of $p_d$ within the domain,
there are two branches, 
where one is denoted as thick (red) line
	 and the other as thin (blue) line
in both (a) and (b).
 }
 \label{fig:1d_p1gauge_evolution}
 \end{figure}

\fig{fig:1d_p1gauge_evolution} 
shows one copy of the two-dimensional spacetimes
that is obtained numerically  from the initial condition of \eq{eq:1d}.
It shows how $q_d$ and the `spatial' metric extraced from \eq{eq:gmunu4}
are correlated with the clock variable, $p_d$.
The classical gauge orbit 
obtained with \eq{eq:lapsep1gauge}
intersects with a constant $p_1$ surface twice
within a finite range of $\tau$
considered in the calculation.
This results in two branches of solution for each value of $p_d$.
If the wavefunction for the physical variables are constructed conditionally 
on the outcome of a measurement of the clock variable\cite{PhysRevD.27.2885},
the physical variables at a fixed $p_d$ are in a linear superposition
of macroscopically distinct states.
The first branch is denoted as the thick (red) line,
while the second branch as the thin (blue) line
in \fig{fig:1d_p1gauge_evolution}. 
Near $p_d \approx 0$ in the first branch,
the space has $+$ signature,
which gives rise to a two-dimensional Lorentzian manifold
(in \eq{gmunu2}, the signature of time is chosen to be $-$ as a convention).
As $p_d$ increases in the first branch,
$g^{11}$ decreases, 
which results in an expanding universe.
At a critical $p_d \approx 0.68$,
the spacetime undergoes a phase transition 
that causes $g^{11}$ to vanish. 
This is a Lifshitz transition 
where the second derivative of 
$U_{k} = \sum_{i,j} e^{i k (r_i - r_j)} U^{ij}$
with respect to $k$ vanishes at zero momentum\footnote{
According to \eq{gmunu2}, the contravariant metric 
is given by
the second moment of $\hat C^{iikkn}_m$.
In the presence of the translational invariance,
the uniform metric can be written as
$g^{\mu \nu}  =   4 \tilde \alpha \left( 
\frac{ \partial U_{\bk}}{\partial k_\mu}  \frac{ \partial U_{k}}{\partial k_\nu} 
+ U_{k}  \frac{ \partial^2 U_{k}}{\partial k_\mu \partial k_\nu}
\right)_{k=0}$\cite{Lee:2020aa}.
With the reflection symmetry, 
$\left. \frac{ \partial U_{\bk}}{\partial k_\mu}  
\right|_{k=0}=0$, 
and the metric is given by the second derivative of $U_k$.
}.
Across the critical point, $g^{11}$ changes the sign,
and the spacetime becomes Euclidean.
At a later time ($p_d \approx 1.34$), 
a second Lifshitz transition restores the Lorentzian signature.
After the second Lifshitz transition, the space shrinks with increasing $p_d$
until it hits `the end of time' around $p_d \approx 2.15$.
At the end point, the first branch converges with the second branch.
In the second branch, the two-dimensional spacetime 
stays as a Lorentzian manifold throughout the evolution.

\subsubsection{Finger 2}
\label{subsubsection:finger2}

 \begin{figure}[ht]
 \begin{center}
   \subfigure[]{
 \includegraphics[scale=0.8]{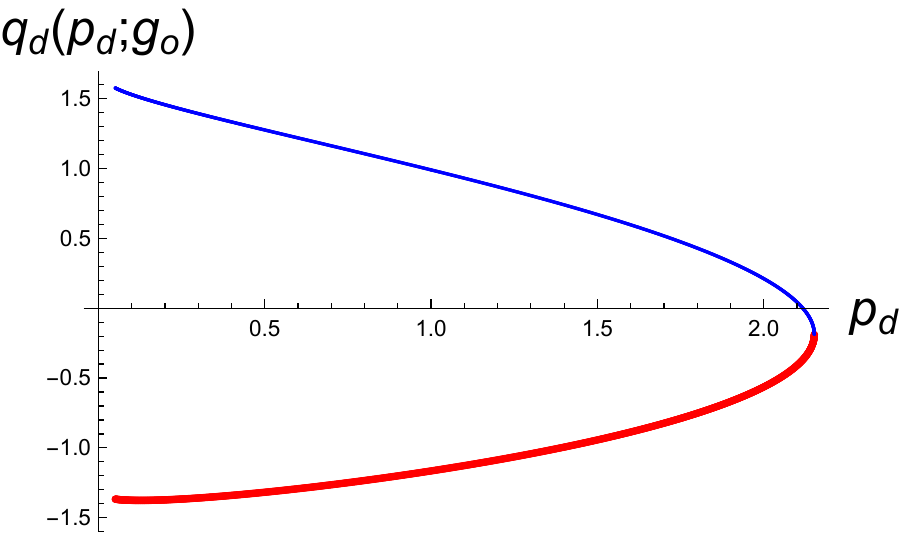} 
  \label{fig:q1p2d}
 } 
  \hfill
 \subfigure[]{
 \includegraphics[scale=0.8]{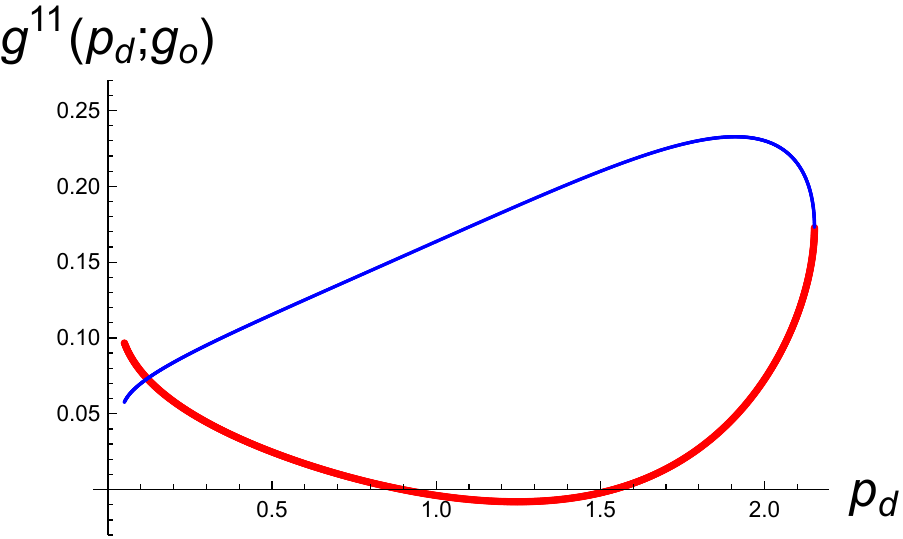} 
 \label{fig:g11p1_2d}
 } 
  \subfigure[]{
 \includegraphics[scale=0.8]{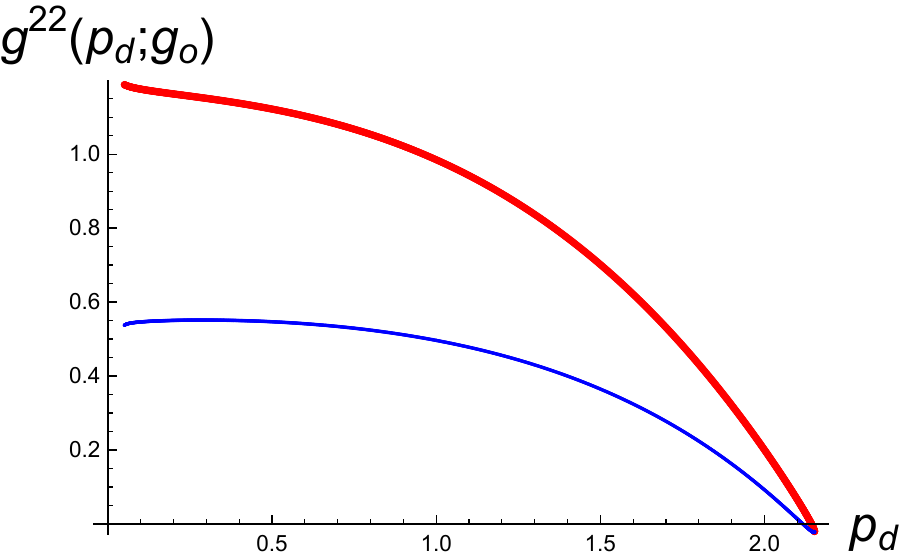} 
 \label{fig:g11p1_2d}
 } 
   \hfill
 \subfigure[]{
 \includegraphics[scale=0.8]{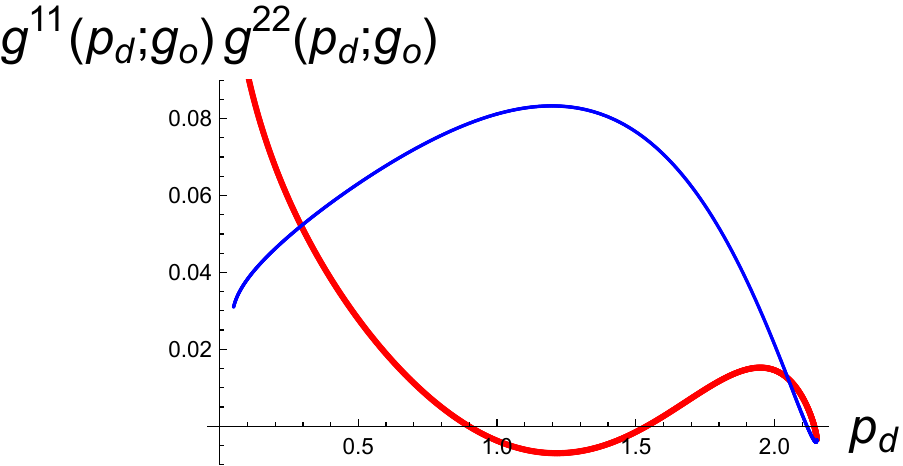} 
 \label{fig:g1and2p1_2d}
 } 
  \end{center}
 \caption{
 The evolution of $q_d$ and two components of the metric
 as functions of $p_d$ 
	 that emerges in the $g_o$-frame with $\zeta=0.1$
 for the same initial condition
 used in \fig{fig:1d_p1gauge_evolution}.
For each value of $p_d$, there exist two solutions.
The first branch is denoted as thick (red) line,
and the second branch as thin (blue) line
in all plots.
 }
 \label{fig:2d_p1gauge_evolution}
 \end{figure}

Now, let us describe the spacetime that emerges 
from the same state in \eq{eq:1d} 
for observers who use a different set of local clocks.
The new local clocks are the diagonal elements of $p_1$
in a different frame.
For concreteness, let us choose clocks that are local in $g_o$-frame,
where 
\bqa
(g_o)^i_j = A_\zeta
\delta_{i_x j_x} 
 \left[
\delta_{i_y j_y} 
+ 
\zeta
\left( 
\delta_{ (i_y-j_y)_{\sL},1} 
+\delta_{ (i_y-j_y)_{\sL},-1} 
\right)
\right].
\label{eq:gstarij}
\eqa
Here $\zeta$ is a constant,
and $A_\zeta = \left[
\prod_{n=1}^{\sqrt{L}} 
 \left(
1
+ 2 \zeta \cos \left(
\frac{2 \pi n}{\sqrt{L}}
\right)
\right)
\right]^{-\frac{1}{\sqrt{L}}}
$.
The new gauge fixing conditions read
\bqa
q = q_d ~ g_o, ~~~ p_1 = p_d ~I.
\label{eq:p2gauge2}
\eqa

The new choice of clocks leads to a different decomposition of the kinematic
Hilbert space into local Hilbert spaces. 
In order to extract the spacetime
that emerges in this new frame,
we need to apply gauge transformations to \eq{eq:1d} 
to enforce the gauge fixing conditions in \eq{eq:p2gauge2}.
As explained in Secs. \ref{subsec:fixshift}
and
\ref{subsec:fixlapse}, 
this is done in two stages.
First, we apply an \SLL transformation
to enforce the first gauge fixing condition, $q = q_d g_o$.
Under the \SLL transformation that brings
$q$ into the form in \eq{eq:p2gauge2},
the collective variables become
\bqa
p_{2}(\tau_1) = (g_o)^T
p_{2} g_o, 
&~~~&
\tilde t_{2}(\tau_1) = 0, \nn
p_{1}(\tau_1) = 
 (g_o)^T p_{1} g_o,
&~~~&
 q(\tau_1) = q_d g_o.
\label{eq:2d2}
\eqa
A site with coordinate $(i_x,i_y)$
in the $g_o$-frame is composed of a linear superposition of 
sites with $(i_x, i_y-1),(i_x, i_y), (i_x,i_y+1)$ in the $I$-frame.
Because one site in the $g_o$-frame  is delocalized across three
neighbouring chains of the $I$-frame,
the chains are no longer decoupled in the $g_o$-frame. 
Due to the interchain entanglement, 
\eq{eq:2d2} has a two-dimensional local structure,
as is shown in \fig{fig:2b}.
Now, we apply the second set of gauge transformations
to enforce the gauge fixing condition for $p_1$.
As explained in Sec. \ref{subsec:fixlapse}, 
this is achieved with the lapse tensor that satisfies 
\eq{eq:eqforv} and the shift tensor given in \eq{eq:shift}.
To understand the nature of this second gauge transformation,
we use the two-dimensional coordinate system, $r_i=(i_x,i_y)$.
This coordinate system makes the two-dimensional local structure manifest. 
In other words, $p_{c,ij}, \td t_c^{ij}$ and $\left(s (q^{-1})^T \right)^{ij}$
connect a site with its neighbours in the two-dimensional manifold,
and decay exponentially in $r_i-r_j$.
As a result, the lapse tensor $v_{ij}$ that satisfies \eq{eq:eqforv} 
also decays exponentially  in $r_i-r_j$.
This implies that the Hamiltonian acts as a
two-dimensional local Hamiltonian along the
gauge orbit that connects 
\eq{eq:2d2} with the one that satisfies
the gauge fixing condition in \eq{eq:p2gauge2}.
Therefore, the  state obtained at the end of the second gauge transformation at $\tau_2$
also supports a two-dimensional local structure in the $g_o$-frame.
The physical variables ($q_d, p_2, \td t_2$) at $\tau_2$ 
viewed as functions of $p_d$
describe a three-dimensional spacetime.

\fig{fig:2d_p1gauge_evolution} shows the evolution of $q_d$,
$g^{11}$ and $g^{22}$ as measured against the physical time $p_d$.
The trajectory is obtained numerically using the same initial condition 
used in \fig{fig:1d_p1gauge_evolution}. 
Due to the reflection symmetry in each direction in space, $g^{12}=0$.
The signature of the three-dimensional spacetime is 
give by $\Bigl( -, \sgn{g^{11}},   \sgn{g^{22}} \Bigr)$.
For each choice of $p_d$, there are two branches of solutions
(the first denoted as thick (red) line and the second denoted
as thin (blue) line).
The evolution of $q_d$ and $g^{11}$ is more or less the same
as the one obtained in the $I$-frame.
What is new here is that $g^{22}$ is non-zero
and exhibits a non-trivial dynamics 
because the state has the two-dimensional local structure.
For each $g^{11}$ and $g^{22}$, 
two Lifshitz transitions occur 
that flip the signature 
of each `spatial' direction
from $+$ to $-$ and back to $+$.
Because the Lifshitz transitions in the $1$ and $2$ directions happen 
at different moments of time,
the signature of the spacetime evolves as 
$(-,+,+) \rightarrow  (-,-,+) \rightarrow  (-,+,+) \rightarrow  (-,+,-) \rightarrow (-,+,+) $ 
as we start from $p_d=0$ in the first branch,
move along the direction of increasing $p_d$
and continue on the second branch.
In  \fig{fig:g1and2p1_2d},
the intervals with $g^{11} g^{22} < 0$ correspond
to the spacetime with two time directions. 
This can be generalized to higher dimensions,
and we expect that anisotropic spacetimes close Lifshitz transitions 
generically exhibit multiple time directions\cite{Bars_2001}.

This example shows that one state can exhibit spacetime manifolds 
with different dimensions, signatures, topologies and geometries in different frames.
This is possible because the enlarged gauge symmetry
generated by \SLL can not only permute sites
but also change the very notion of local sites 
by constructing new sites out of linear superpositions of old sites.

If one chooses local clocks in an arbitrary frame,  
the state generally does not retain any local structure.
Even for a state that has a local structure in one frame,
a well-defined spacetime manifold does not emerge
if a collection of clocks are chosen
in another frame that is related to the first frame through
a non-local transformation\footnote{
Equivalently, a state that is short-range entangled in one basis
can exhibit long-range entanglement 
if one chooses non-local basis.
}.
The emergence of a well-defined spacetime
hinges both on local structure of the state
and on the choice of local clocks
that are compatible with the local structure of the state.

\section{Summary and discussion}
\label{sec:summary}

In this paper, we consider a theory of quantum gravity
that does not have a preferred decomposition
of the kinematic Hilbert space into local Hilbert spaces.
The theory is covariant under a gauge symmetry larger than diffeomorphism,
where the extra gauge symmetry includes transformations 
that mix local kinematic Hilbert spaces.
This gives rise to a greater freedom in choosing a collection of local clocks
with respect to which the evolution of other physical degrees of freedom
is tracked.
It is shown that dimension, signature, topology and geometry
of spacetime depend on the choice of local clocks. 
Just as a gem reveals different facets in different cuts,
one state can exhibit different spacetimes 
with different choices of clocks.
We expect that this is a generic feature
of theories that do not have 
a preferred Hilbert space decomposition.

Another consequence of the enlarged gauge symmetry
is the presence of extra propagating modes 
besides the spin $2$ gravitational mode.
They are represented by the higher-spin fields 
associated with the bi-local collective fields.
Higher-spin gauge fields
are Higgsed in states 
that break \SLL to the global translation symmetry,
as is the case for the states considered in 
Sec. \ref{subsec:more}\cite{Lee:2020aa}.
It will be of interest to understand the physical spectrum of the theory.

\section*{Acknowledgments}

The research was supported by
the Natural Sciences and Engineering Research Council of
Canada.
Research at Perimeter Institute is supported in part by the Government of Canada through the Department of Innovation, Science and Economic Development Canada and by the Province of Ontario through the Ministry of Colleges and Universities.

 \bibliography{references}

\begin{thebibliography}{30}
\expandafter\ifx\csname natexlab\endcsname\relax\def\natexlab#1{#1}\fi
\expandafter\ifx\csname bibnamefont\endcsname\relax
  \def\bibnamefont#1{#1}\fi
\expandafter\ifx\csname bibfnamefont\endcsname\relax
  \def\bibfnamefont#1{#1}\fi
\expandafter\ifx\csname citenamefont\endcsname\relax
  \def\citenamefont#1{#1}\fi
\expandafter\ifx\csname url\endcsname\relax
  \def\url#1{\texttt{#1}}\fi
\expandafter\ifx\csname urlprefix\endcsname\relax\def\urlprefix{URL }\fi
\providecommand{\bibinfo}[2]{#2}
\providecommand{\eprint}[2][]{\url{#2}}

\bibitem[{\citenamefont{DeWitt}(1967)}]{PhysRev.160.1113}
\bibinfo{author}{\bibfnamefont{B.~S.} \bibnamefont{DeWitt}},
  \bibinfo{journal}{Phys. Rev.} \textbf{\bibinfo{volume}{160}},
  \bibinfo{pages}{1113} (\bibinfo{year}{1967}),
  \urlprefix\url{https://link.aps.org/doi/10.1103/PhysRev.160.1113}.

\bibitem[{\citenamefont{{Isham}}(1992)}]{1992gr.qc....10011I}
\bibinfo{author}{\bibfnamefont{C.~J.} \bibnamefont{{Isham}}},
  \bibinfo{journal}{ArXiv General Relativity and Quantum Cosmology e-prints}
  (\bibinfo{year}{1992}), \eprint{gr-qc/9210011}.

\bibitem[{\citenamefont{{Kuchar}}(1992)}]{1992grra.conf..211K}
\bibinfo{author}{\bibfnamefont{K.~V.} \bibnamefont{{Kuchar}}}, in
  \emph{\bibinfo{booktitle}{4th Canadian Conference on General Relativity and
  Relativistic Astrophysics}}, edited by
  \bibinfo{editor}{\bibfnamefont{G.}~\bibnamefont{{Kunstatter}}},
  \bibinfo{editor}{\bibfnamefont{D.~E.} \bibnamefont{{Vincent}}},
  \bibnamefont{and} \bibinfo{editor}{\bibfnamefont{J.~G.}
  \bibnamefont{{Williams}}} (\bibinfo{year}{1992}), p. \bibinfo{pages}{211}.

\bibitem[{\citenamefont{Anderson}(2012)}]{https://doi.org/10.1002/andp.201200147}
\bibinfo{author}{\bibfnamefont{E.}~\bibnamefont{Anderson}},
  \bibinfo{journal}{Annalen der Physik} \textbf{\bibinfo{volume}{524}},
  \bibinfo{pages}{757} (\bibinfo{year}{2012}),
  \eprint{https://onlinelibrary.wiley.com/doi/pdf/10.1002/andp.201200147},
  \urlprefix\url{https://onlinelibrary.wiley.com/doi/abs/10.1002/andp.201200147}.

\bibitem[{\citenamefont{Rovelli}(2002)}]{PhysRevD.65.124013}
\bibinfo{author}{\bibfnamefont{C.}~\bibnamefont{Rovelli}},
  \bibinfo{journal}{Phys. Rev. D} \textbf{\bibinfo{volume}{65}},
  \bibinfo{pages}{124013} (\bibinfo{year}{2002}),
  \urlprefix\url{https://link.aps.org/doi/10.1103/PhysRevD.65.124013}.

\bibitem[{\citenamefont{Dittrich}(2007)}]{Dittrich2007}
\bibinfo{author}{\bibfnamefont{B.}~\bibnamefont{Dittrich}},
  \bibinfo{journal}{General Relativity and Gravitation}
  \textbf{\bibinfo{volume}{39}}, \bibinfo{pages}{1891} (\bibinfo{year}{2007}),
  ISSN \bibinfo{issn}{1572-9532},
  \urlprefix\url{https://doi.org/10.1007/s10714-007-0495-2}.

\bibitem[{\citenamefont{Gambini et~al.}(2004)\citenamefont{Gambini, Porto, and
  Pullin}}]{Gambini_2004}
\bibinfo{author}{\bibfnamefont{R.}~\bibnamefont{Gambini}},
  \bibinfo{author}{\bibfnamefont{R.~A.} \bibnamefont{Porto}}, \bibnamefont{and}
  \bibinfo{author}{\bibfnamefont{J.}~\bibnamefont{Pullin}},
  \bibinfo{journal}{New Journal of Physics} \textbf{\bibinfo{volume}{6}},
  \bibinfo{pages}{45} (\bibinfo{year}{2004}),
  \urlprefix\url{https://doi.org/10.1088/1367-2630/6/1/045}.

\bibitem[{\citenamefont{Lee}(2020)}]{Lee:2020aa}
\bibinfo{author}{\bibfnamefont{S.-S.} \bibnamefont{Lee}},
  \bibinfo{journal}{Journal of High Energy Physics}
  \textbf{\bibinfo{volume}{2020}}, \bibinfo{pages}{70} (\bibinfo{year}{2020}),
  \urlprefix\url{https://doi.org/10.1007/JHEP06(2020)070}.

\bibitem[{\citenamefont{Maldacena}(1999)}]{Maldacena:1997re}
\bibinfo{author}{\bibfnamefont{J.~M.} \bibnamefont{Maldacena}},
  \bibinfo{journal}{Int.J.Theor.Phys.} \textbf{\bibinfo{volume}{38}},
  \bibinfo{pages}{1113} (\bibinfo{year}{1999}), \eprint{hep-th/9711200}.

\bibitem[{\citenamefont{Witten}(1998)}]{Witten:1998qj}
\bibinfo{author}{\bibfnamefont{E.}~\bibnamefont{Witten}},
  \bibinfo{journal}{Adv.Theor.Math.Phys.} \textbf{\bibinfo{volume}{2}},
  \bibinfo{pages}{253} (\bibinfo{year}{1998}), \eprint{hep-th/9802150}.

\bibitem[{\citenamefont{Gubser et~al.}(1998)\citenamefont{Gubser, Klebanov, and
  Polyakov}}]{Gubser:1998bc}
\bibinfo{author}{\bibfnamefont{S.}~\bibnamefont{Gubser}},
  \bibinfo{author}{\bibfnamefont{I.~R.} \bibnamefont{Klebanov}},
  \bibnamefont{and} \bibinfo{author}{\bibfnamefont{A.~M.}
  \bibnamefont{Polyakov}}, \bibinfo{journal}{Phys.Lett.}
  \textbf{\bibinfo{volume}{B428}}, \bibinfo{pages}{105} (\bibinfo{year}{1998}),
  \eprint{hep-th/9802109}.

\bibitem[{\citenamefont{Arnowitt et~al.}(1959)\citenamefont{Arnowitt, Deser,
  and Misner}}]{PhysRev.116.1322}
\bibinfo{author}{\bibfnamefont{R.}~\bibnamefont{Arnowitt}},
  \bibinfo{author}{\bibfnamefont{S.}~\bibnamefont{Deser}}, \bibnamefont{and}
  \bibinfo{author}{\bibfnamefont{C.~W.} \bibnamefont{Misner}},
  \bibinfo{journal}{Phys. Rev.} \textbf{\bibinfo{volume}{116}},
  \bibinfo{pages}{1322} (\bibinfo{year}{1959}),
  \urlprefix\url{https://link.aps.org/doi/10.1103/PhysRev.116.1322}.

\bibitem[{\citenamefont{Teitelboim}(1973)}]{TEITELBOIM1973542}
\bibinfo{author}{\bibfnamefont{C.}~\bibnamefont{Teitelboim}},
  \bibinfo{journal}{Annals of Physics} \textbf{\bibinfo{volume}{79}},
  \bibinfo{pages}{542 } (\bibinfo{year}{1973}), ISSN \bibinfo{issn}{0003-4916},
  \urlprefix\url{http://www.sciencedirect.com/science/article/pii/0003491673900961}.

\bibitem[{\citenamefont{Ryu and Takayanagi}(2006)}]{PhysRevLett.96.181602}
\bibinfo{author}{\bibfnamefont{S.}~\bibnamefont{Ryu}} \bibnamefont{and}
  \bibinfo{author}{\bibfnamefont{T.}~\bibnamefont{Takayanagi}},
  \bibinfo{journal}{Phys. Rev. Lett.} \textbf{\bibinfo{volume}{96}},
  \bibinfo{pages}{181602} (\bibinfo{year}{2006}),
  \urlprefix\url{http://link.aps.org/doi/10.1103/PhysRevLett.96.181602}.

\bibitem[{\citenamefont{Hubeny et~al.}(2007)\citenamefont{Hubeny, Rangamani,
  and Takayanagi}}]{1126-6708-2007-07-062}
\bibinfo{author}{\bibfnamefont{V.~E.} \bibnamefont{Hubeny}},
  \bibinfo{author}{\bibfnamefont{M.}~\bibnamefont{Rangamani}},
  \bibnamefont{and}
  \bibinfo{author}{\bibfnamefont{T.}~\bibnamefont{Takayanagi}},
  \bibinfo{journal}{Journal of High Energy Physics}
  \textbf{\bibinfo{volume}{2007}}, \bibinfo{pages}{062} (\bibinfo{year}{2007}),
  \urlprefix\url{http://stacks.iop.org/1126-6708/2007/i=07/a=062}.

\bibitem[{\citenamefont{Van~Raamsdonk}(2010)}]{VanRaamsdonk:2010pw}
\bibinfo{author}{\bibfnamefont{M.}~\bibnamefont{Van~Raamsdonk}},
  \bibinfo{journal}{Gen. Rel. Grav.} \textbf{\bibinfo{volume}{42}},
  \bibinfo{pages}{2323} (\bibinfo{year}{2010}), \bibinfo{note}{[Int. J. Mod.
  Phys.D19,2429(2010)]}, \eprint{1005.3035}.

\bibitem[{\citenamefont{Lewkowycz and Maldacena}(2013)}]{Lewkowycz2013}
\bibinfo{author}{\bibfnamefont{A.}~\bibnamefont{Lewkowycz}} \bibnamefont{and}
  \bibinfo{author}{\bibfnamefont{J.}~\bibnamefont{Maldacena}},
  \bibinfo{journal}{Journal of High Energy Physics}
  \textbf{\bibinfo{volume}{2013}}, \bibinfo{pages}{1} (\bibinfo{year}{2013}),
  ISSN \bibinfo{issn}{1029-8479},
  \urlprefix\url{http://dx.doi.org/10.1007/JHEP08(2013)090}.

\bibitem[{\citenamefont{Headrick et~al.}(2014)\citenamefont{Headrick, Hubeny,
  Lawrence, and Rangamani}}]{Headrick:2014aa}
\bibinfo{author}{\bibfnamefont{M.}~\bibnamefont{Headrick}},
  \bibinfo{author}{\bibfnamefont{V.~E.} \bibnamefont{Hubeny}},
  \bibinfo{author}{\bibfnamefont{A.}~\bibnamefont{Lawrence}}, \bibnamefont{and}
  \bibinfo{author}{\bibfnamefont{M.}~\bibnamefont{Rangamani}},
  \bibinfo{journal}{Journal of High Energy Physics}
  \textbf{\bibinfo{volume}{2014}}, \bibinfo{pages}{162} (\bibinfo{year}{2014}),
  \urlprefix\url{https://doi.org/10.1007/JHEP12(2014)162}.

\bibitem[{\citenamefont{Faulkner et~al.}(2013)\citenamefont{Faulkner,
  Lewkowycz, and Maldacena}}]{Faulkner:2013aa}
\bibinfo{author}{\bibfnamefont{T.}~\bibnamefont{Faulkner}},
  \bibinfo{author}{\bibfnamefont{A.}~\bibnamefont{Lewkowycz}},
  \bibnamefont{and}
  \bibinfo{author}{\bibfnamefont{J.}~\bibnamefont{Maldacena}},
  \bibinfo{journal}{Journal of High Energy Physics}
  \textbf{\bibinfo{volume}{2013}}, \bibinfo{pages}{74} (\bibinfo{year}{2013}),
  \urlprefix\url{https://doi.org/10.1007/JHEP11(2013)074}.

\bibitem[{\citenamefont{Lashkari et~al.}(2014)\citenamefont{Lashkari,
  McDermott, and Van~Raamsdonk}}]{Lashkari:2014aa}
\bibinfo{author}{\bibfnamefont{N.}~\bibnamefont{Lashkari}},
  \bibinfo{author}{\bibfnamefont{M.~B.} \bibnamefont{McDermott}},
  \bibnamefont{and}
  \bibinfo{author}{\bibfnamefont{M.}~\bibnamefont{Van~Raamsdonk}},
  \bibinfo{journal}{Journal of High Energy Physics}
  \textbf{\bibinfo{volume}{2014}}, \bibinfo{pages}{195} (\bibinfo{year}{2014}),
  \urlprefix\url{https://doi.org/10.1007/JHEP04(2014)195}.

\bibitem[{\citenamefont{{Qi}}(2013)}]{2013arXiv1309.6282Q}
\bibinfo{author}{\bibfnamefont{X.-L.} \bibnamefont{{Qi}}},
  \bibinfo{journal}{ArXiv e-prints}  (\bibinfo{year}{2013}),
  \eprint{1309.6282}.

\bibitem[{\citenamefont{{Faulkner} et~al.}(2014)\citenamefont{{Faulkner},
  {Guica}, {Hartman}, {Myers}, and {Van Raamsdonk}}}]{2014JHEP...03..051F}
\bibinfo{author}{\bibfnamefont{T.}~\bibnamefont{{Faulkner}}},
  \bibinfo{author}{\bibfnamefont{M.}~\bibnamefont{{Guica}}},
  \bibinfo{author}{\bibfnamefont{T.}~\bibnamefont{{Hartman}}},
  \bibinfo{author}{\bibfnamefont{R.~C.} \bibnamefont{{Myers}}},
  \bibnamefont{and} \bibinfo{author}{\bibfnamefont{M.}~\bibnamefont{{Van
  Raamsdonk}}}, \bibinfo{journal}{Journal of High Energy Physics}
  \textbf{\bibinfo{volume}{3}}, \bibinfo{eid}{51} (\bibinfo{year}{2014}),
  \eprint{1312.7856}.

\bibitem[{\citenamefont{Cao et~al.}(2017)\citenamefont{Cao, Carroll, and
  Michalakis}}]{PhysRevD.95.024031}
\bibinfo{author}{\bibfnamefont{C.}~\bibnamefont{Cao}},
  \bibinfo{author}{\bibfnamefont{S.~M.} \bibnamefont{Carroll}},
  \bibnamefont{and}
  \bibinfo{author}{\bibfnamefont{S.}~\bibnamefont{Michalakis}},
  \bibinfo{journal}{Phys. Rev. D} \textbf{\bibinfo{volume}{95}},
  \bibinfo{pages}{024031} (\bibinfo{year}{2017}),
  \urlprefix\url{https://link.aps.org/doi/10.1103/PhysRevD.95.024031}.

\bibitem[{\citenamefont{Maldacena and
  Susskind}(2013)}]{doi:10.1002/prop.201300020}
\bibinfo{author}{\bibfnamefont{J.}~\bibnamefont{Maldacena}} \bibnamefont{and}
  \bibinfo{author}{\bibfnamefont{L.}~\bibnamefont{Susskind}},
  \bibinfo{journal}{Fortschritte der Physik} \textbf{\bibinfo{volume}{61}},
  \bibinfo{pages}{781} (\bibinfo{year}{2013}),
  \eprint{https://onlinelibrary.wiley.com/doi/pdf/10.1002/prop.201300020},
  \urlprefix\url{https://onlinelibrary.wiley.com/doi/abs/10.1002/prop.201300020}.

\bibitem[{\citenamefont{Lee}(2018)}]{Lee2018}
\bibinfo{author}{\bibfnamefont{S.-S.} \bibnamefont{Lee}},
  \bibinfo{journal}{Journal of High Energy Physics}
  \textbf{\bibinfo{volume}{2018}}, \bibinfo{pages}{43} (\bibinfo{year}{2018}),
  ISSN \bibinfo{issn}{1029-8479},
  \urlprefix\url{https://doi.org/10.1007/JHEP10(2018)043}.

\bibitem[{\citenamefont{Lee}(2019)}]{Lee2019}
\bibinfo{author}{\bibfnamefont{S.-S.} \bibnamefont{Lee}},
  \bibinfo{journal}{Journal of High Energy Physics}
  \textbf{\bibinfo{volume}{2019}}, \bibinfo{pages}{215} (\bibinfo{year}{2019}),
  ISSN \bibinfo{issn}{1029-8479},
  \urlprefix\url{https://doi.org/10.1007/JHEP05(2019)215}.

\bibitem[{\citenamefont{Lee}(2016)}]{Lee2016}
\bibinfo{author}{\bibfnamefont{S.-S.} \bibnamefont{Lee}},
  \bibinfo{journal}{Journal of High Energy Physics}
  \textbf{\bibinfo{volume}{2016}}, \bibinfo{pages}{44} (\bibinfo{year}{2016}),
  ISSN \bibinfo{issn}{1029-8479},
  \urlprefix\url{https://doi.org/10.1007/JHEP09(2016)044}.

\bibitem[{\citenamefont{Page and Wootters}(1983)}]{PhysRevD.27.2885}
\bibinfo{author}{\bibfnamefont{D.~N.} \bibnamefont{Page}} \bibnamefont{and}
  \bibinfo{author}{\bibfnamefont{W.~K.} \bibnamefont{Wootters}},
  \bibinfo{journal}{Phys. Rev. D} \textbf{\bibinfo{volume}{27}},
  \bibinfo{pages}{2885} (\bibinfo{year}{1983}),
  \urlprefix\url{https://link.aps.org/doi/10.1103/PhysRevD.27.2885}.

\bibitem[{\citenamefont{Bars}(2001)}]{Bars_2001}
\bibinfo{author}{\bibfnamefont{I.}~\bibnamefont{Bars}},
  \bibinfo{journal}{Classical and Quantum Gravity}
  \textbf{\bibinfo{volume}{18}}, \bibinfo{pages}{3113} (\bibinfo{year}{2001}),
  \urlprefix\url{https://doi.org/10.1088%2F0264-9381%2F18%2F16%2F303}.

\bibitem[{\citenamefont{Amelino-Camelia
  et~al.}(2011)\citenamefont{Amelino-Camelia, Freidel, Kowalski-Glikman, and
  Smolin}}]{PhysRevD.84.084010}
\bibinfo{author}{\bibfnamefont{G.}~\bibnamefont{Amelino-Camelia}},
  \bibinfo{author}{\bibfnamefont{L.}~\bibnamefont{Freidel}},
  \bibinfo{author}{\bibfnamefont{J.}~\bibnamefont{Kowalski-Glikman}},
  \bibnamefont{and} \bibinfo{author}{\bibfnamefont{L.}~\bibnamefont{Smolin}},
  \bibinfo{journal}{Phys. Rev. D} \textbf{\bibinfo{volume}{84}},
  \bibinfo{pages}{084010} (\bibinfo{year}{2011}),
  \urlprefix\url{https://link.aps.org/doi/10.1103/PhysRevD.84.084010}.

\end{thebibliography}

 \newpage
 \appendix
 \section{Gauge invariance of \eq{eq:project} }
 \label{eq:appgaugeinv}

 Here we prove that \eq{eq:project} is gauge invariant.
We write the set of constraints as 
  $\{ \hat C_1, \hat C_2, ..., \hat C_n \}$,
  where each element represents one component
  of $\hG^i_{~j}$ or $\hH^{kl}$.
  Because $\hG$ is traceless and $\hH$ is symmetric,
$n = L^2 - 1 + \frac{L(L+1)}{2}$.
The associated shift and lapse tensors 
are written as $n$ gauge parameters,  
$ x = (x_1,x_2,..,x_n) \in \mathbb{R}^n$.
\eq{eq:project} can be written as
 \bqa
 \cb 0_\chi \rb 
=
\int_{\bf x \in A} \cD {\bf x}~ U( {\bf x} )
\cb 0 \rb.
\label{eq:0chiapp}
\eqa
Here,
$ U( {\bf x} ) = 
\bar {\cal P}_l \left[
\prod_{l=1}^\infty
e^{-i  \hat C \cdot x^{(l)}  }
\right] 
$
with
 $ \hat C \cdot  x \equiv \sum_{i=1}^n \hat C_i x_i$.
In the definition of $U({\bf x})$,
$ \bar {\cal P}_l$ orders the unitary operators so that 
$e^{-i  \hat C \cdot x^{(l)}  }$ with smaller $l$ are placed 
to the left of the terms with larger $l$. 
 $\cD {\bf x} \equiv \int \prod_{l=1}^\infty Dx^{(l)}$,
 and
 $A$ is a set of ordered gauge parameters
$A = \Bigl\{ (x^{(1)},  x^{(2)}, ... ) \cb  x^{(l)} \in \mathbb{R}^n, l=1,2,..,\infty \Bigr\}$.
 $\varepsilon$ in \eq{eq:project} has been absorbed into the gauge parameter.
Now, we consider a state obtained by applying a gauge transformation
to \eq{eq:0chiapp}, 
 \bqa
 \cb 0'_\chi \rb 
=
e^{-i  \hat C \cdot \td x  \ }  \cb 0_\chi \rb 
\label{eq:0primechi}
 \eqa
for $\td x \in \mathbb{R}^n$.
\eq{eq:0primechi} can be written as
 \bqa
 \cb 0_\chi' \rb 
=
\int_{\bf x \in A'} D {\bf x}~ U( {\bf x} )
\cb 0 \rb,
\label{eq:0chiapp2}
\eqa
    where 
  $  A' = \Bigl\{ (  \td x,  x^{(1)},  x^{(2)}, ... ) \cb  x^{(l)} \in \mathbb{R}^n, l=1,2,..,\infty \Bigr\}$.
Now we prove that
$W = \Bigl\{  U({\bf x})  \cb 
    {\bf x } \in A \Bigr\}$
    and 
$W' = \Bigl\{  U({\bf x}')  \cb 
    {\bf x }' \in A' \Bigr\}$
 are the same.
For every element $U( {\bf x}') \in W'$,
there exists ${\bf x} = ( \td x, x^{(1)},  x^{(2)}, ... )$ in $A$
such that $U( {\bf x} )  = U( {\bf x}')$.
This shows that $W' \subset W$.
Conversely,
for every element $U( {\bf x}) \in W$,
there exists ${\bf x}' = ( \td x, - \td x, x^{(1)},  x^{(2)}, ... )$ in $A'$
such that $U( {\bf x}' )  = U( {\bf x})$.
This shows that $W \subset W'$.
Therefore, $W = W'$.
If $\cb 0_\chi \rb$ does not vanish, 
$ \cb 0_\chi \rb  =  \cb 0_\chi' \rb $.

\end{document}